\documentclass[useAMS,usenatbib,usegraphicx]{mn2e}
\usepackage{pslatex}
\usepackage{pdfscreen}
\usepackage{amsmath,amsfonts}
\newcommand{\Hsec}{\mathcal{H}_{\idm{sec}}}

\def\ogle{OGLE-2006-BLG-109L}
\def\idm#1{{\mbox{\scriptsize #1}}}
\def\Ym{\left<Y\right>}
\newcommand{\fHz}{f_{\mbox{\scriptsize HZ}}}
\newcommand{\sHz}{s_{\mbox{\scriptsize HZ}}}

\def\hide#1{{}}
\title[Habitable zone of the OGLE-2006-BLG-109L system]
{On the habitability of the OGLE-2006-BLG-109L planetary system}
\author[C. Migaszewski, K. Go\'zdziewski \& T.C. Hinse]{
Cezary Migaszewski$^{1}$\thanks{E-mail: c.migaszewski@astri.umk.pl},
Krzysztof Go\'zdziewski$^{1}$\footnotemark[1]\thanks{E-mail:k.gozdziewski@astri.umk.pl},
and Tobias C. Hinse$^2$\footnotemark[2]\thanks{E-mail: tobiash@arm.ac.uk}\\
$^{1}$Toru\'n Centre for Astronomy, Nicolaus Copernicus University, Gagarin Str. 11, 87-100 Toru\'n,
Poland\\
$^{2}$Armagh Observatory, College Hill, BT61 9DG Armagh, Northern Ireland, UK}
\begin{document}
%
\date{Accepted 2009 January 16.  Received 2009 January 16; in original form 2008
November 10}
\pagerange{\pageref{firstpage}--\pageref{lastpage}} \pubyear{2008}
\maketitle
\label{firstpage}
%
\begin{abstract}
We investigate the dynamics of putative Earth-mass planets in the habitable zone
(HZ) of the extrasolar planetary system \ogle{}, a close analog of the Solar
system.  Our work is inspired by work of Malhotra and Minton (2008). Using the
linear Laplace--Lagrange theory, they identified a strong secular resonance that
may excite large eccentricity of orbits in the HZ. However, due to uncertain or
unconstrained orbital parameters, the sub-system of Jupiters may be found in
dynamically active region of the phase space spanned by low-order mean-motion
resonances. To generalize this secular model, we construct a semi-analytical
averaging method in terms of the restricted problem. The secular orbits of large
planets are approximated by numerically averaged osculating elements. They are
used to calculate the mean orbits of terrestrial planets by means of a
high-order analytic secular theory developed in our previous works. We found
regions in the parameter space of the problem in which stable, quasi-circular
orbits in the HZ are permitted. The excitation of eccentricity in the HZ
strongly depends on the apsidal angle of jovian orbits.  For some combinations
of that angle, eccentricities and semi-major axes consistent with the
observations, a terrestrial planet may survive in low eccentric orbits. We also
study the effect of post-Newtonian gravity correction on the innermost secular
resonance.
\end{abstract}
%
\begin{keywords}
celestial mechanics -- secular dynamics -- relativistic effects -- 
analytical methods --  extrasolar planetary systems
\end{keywords}
%
\section{Introduction}
%
Recently, \cite{Gaudi2008} announced a discovery of two-planet extrasolar system
hosted by the \ogle{} star of $\sim 0.5$~M$_{\sun}$. Its jovian companions of
$\sim 0.71$ and $\sim 0.27$ Jupiter masses are in orbits of $2.3$ and $4.6$~au,
resembling a scaled copy of the Jupiter-Saturn pair. This similarity leads to a
natural question, whether additional planets can exist in this system, in
particular Earth-like planets in the habitable zone (HZ).  As for now, the
answer can be speculative, because the \ogle{} planetary system has been
detected during an observed micro-lensing event and hence the observational data
are not reproducible. In addition, the faint and distant parent star cannot yet
be observed by other planet-hunting techniques, like the Doppler spectroscopy or
astrometry, to eventually detect the presence of an Earth-like planet. Moreover,
some orbital parameters of the jovian planets are either uncertain or
undetermined at all. The orbital architecture of the system, including putative 
terrestrial planets, and its stability,  may be investigated only indirectly by
numerical simulations.

In this work, we focus on the dynamical structure of the HZ. The HZ is defined
implicitly through a requirement of the liquid state of water \citep[see,
e.g.,][]{Kasting1993,Hinse2008,Malhotra2008}. That condition is satisfied if the
terrestrial orbit, parameterized by the semi-major axis $a_0$ and eccentricity
$e_0$, is confined to the annulus of $(r_p,r_a) \equiv  (0.25, 0.36)$~au 
[following \cite{Hinse2008}, the terrestrial planet in the HZ will be called
with Latin word {\em Tellus} from hereafter and we will always refer to its
orbital elements with index ``0'']. It means that the pericenter and apocenter
distances of habitable orbit must be limited through the following conditions :
\begin{equation}
a_0 (1 - e_0) > r_p \qquad \mbox{and} \qquad a_0 (1 + e_0) < r_a,
\label{eq:eq1}
\end{equation}
providing very long time-scale of stable motion, well enough to develop and
support life \citep[see, e.g.][for details]{Hinse2008}. The term {\em stable} or
{\em HZ-stable}, regarding the terrestrial orbit,  will be understood as
``confined to the HZ, as defined above, during the secular time scale counted in
tens of Myrs''. Apparently, the HZ of the \ogle{} system is separated from the 
resonant influence of the primaries. The ratio of orbital periods  corresponding
to  $a_0 \in (0.25,0.36)$~au is $\sim 20$. Therefore, no strong  mean motion
resonances (MMRs) can perturb putative terrestrial orbits\hide{(see also
Sect.~3)}. However, even in the absence of  the MMRs, such orbits may be still
disturbed by the long-term, secular interactions with jovian companions. 
Indeed, \cite{Malhotra2008} identified two secular resonances ($\nu_{1,2}$) in 
the inner part of the \ogle{} system  related to the fundamental secular
frequencies (modes) of the orbits of Jupiters. These resonances are analogs of
the $\nu_{5,6}$ resonances with Jupiter and Saturn in the Solar system. Both of
them lead to strong amplification of $e_0$ to large values (up to 0.8--0.9), and
are centered around $a_0 \sim 0.3$~au, and $a_0 \sim 0.7$~au, respectively. In
particular, the $\nu_1$ resonance acts in the region of HZ, suppressing
long-term stable motion  confined to that zone. The motion of a terrestrial
planet may be stabilized through a change of the secular frequencies by
interactions with an additional (yet unknown or hypothetical) inner planet
\citep{Malhotra2008}. Such an object of $\sim 0.3~M_{\earth}$ and semi-major
axis $\sim 0.1$~au would change the $\nu_1$  secular frequency, and shift the
``troublesome'' resonance out of the HZ. Similarly, numerical studies of
\cite{Pilat2008A,Pilat2008B},  regarding Solar-like systems show that a small
planet in the inner region may change the position of the secular resonance
$\nu_5$ and significantly affect the habitability for a given semi-major axis.
Earlier, the long term effects of adding or removing planets in the inner Solar
system were analyzed by \cite{Namouni1999}, following \cite{Innanen1998} who
found that the Earth-Moon (EM) stabilizes the orbits of Venus and Mercury by
suppressing a strong secular resonance of period 8.1 Myr near Venus orbit.
Actually, \cite{Namouni1999} conclude that is not clear whether the absence of
EM does really provide a physical insight into the stability of the whole
system. They argue that {\em throughout the accretion of planetesimals, the mass
that formed Earth contributed consistently to the secular precession rates and
more generally to the current state of the Solar system}.  Hence, the studies of
long-term behavior of a planetary system with arbitrary initial conditions
(e.g., modified by adding or removing  planets while keeping other orbital
elements fixed), which are partially consistent with observations, seem not well
justified in general.  Hence, because the \ogle{}-system architecture is not
fully determined (in fact,  it is known poorly, as we will see below), rather to
analyse the influence of  a few hypothetical bodies in the system on the
dynamical structure of the HZ, we follow the known observational constraints on
the system. Its Jovian planets are orbitally coupled, and the habitable zone is
very close to the star, prohibiting extensive numerical integrations. Hence, 
the main motivation of our work is to build up and investigate an appropriate
semi-analytic, specific  model of the system, and try to answer whether
putative, single Earth-like planet may survive in the HZ during the secular
time-scale. 
 
\begin{figure*}
\centerline{
  \vbox{
        \hbox{\includegraphics[  width=119mm]{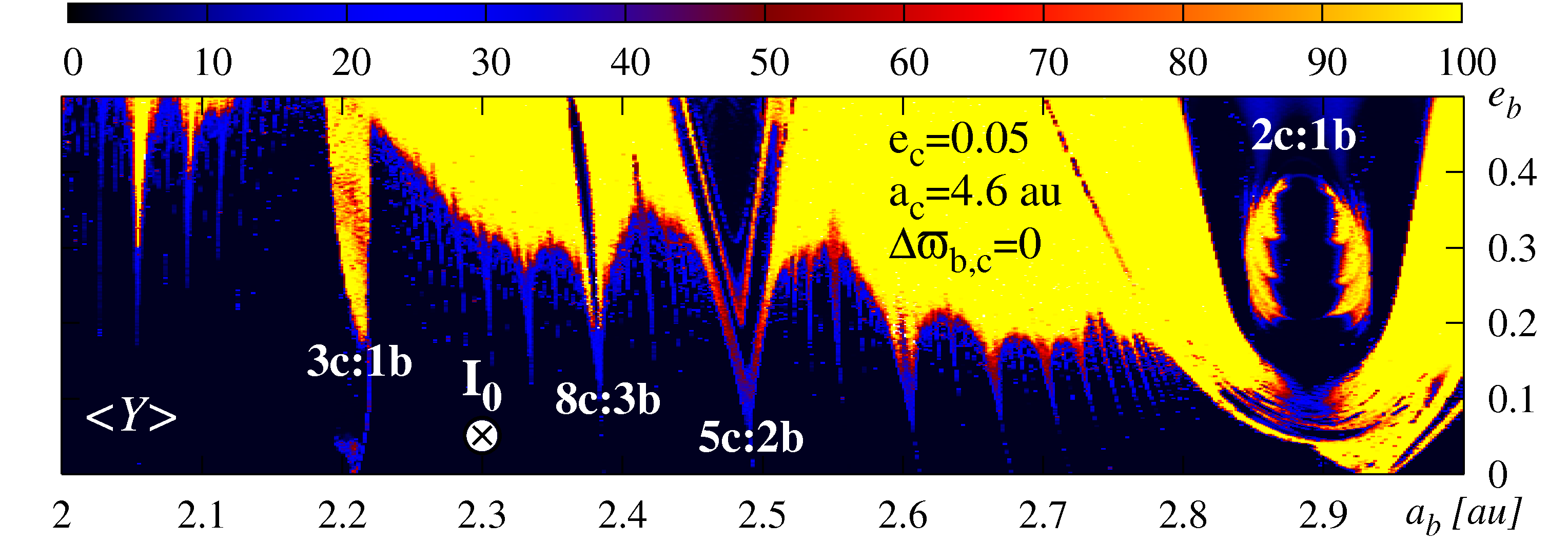}
              \includegraphics[  width=56mm]{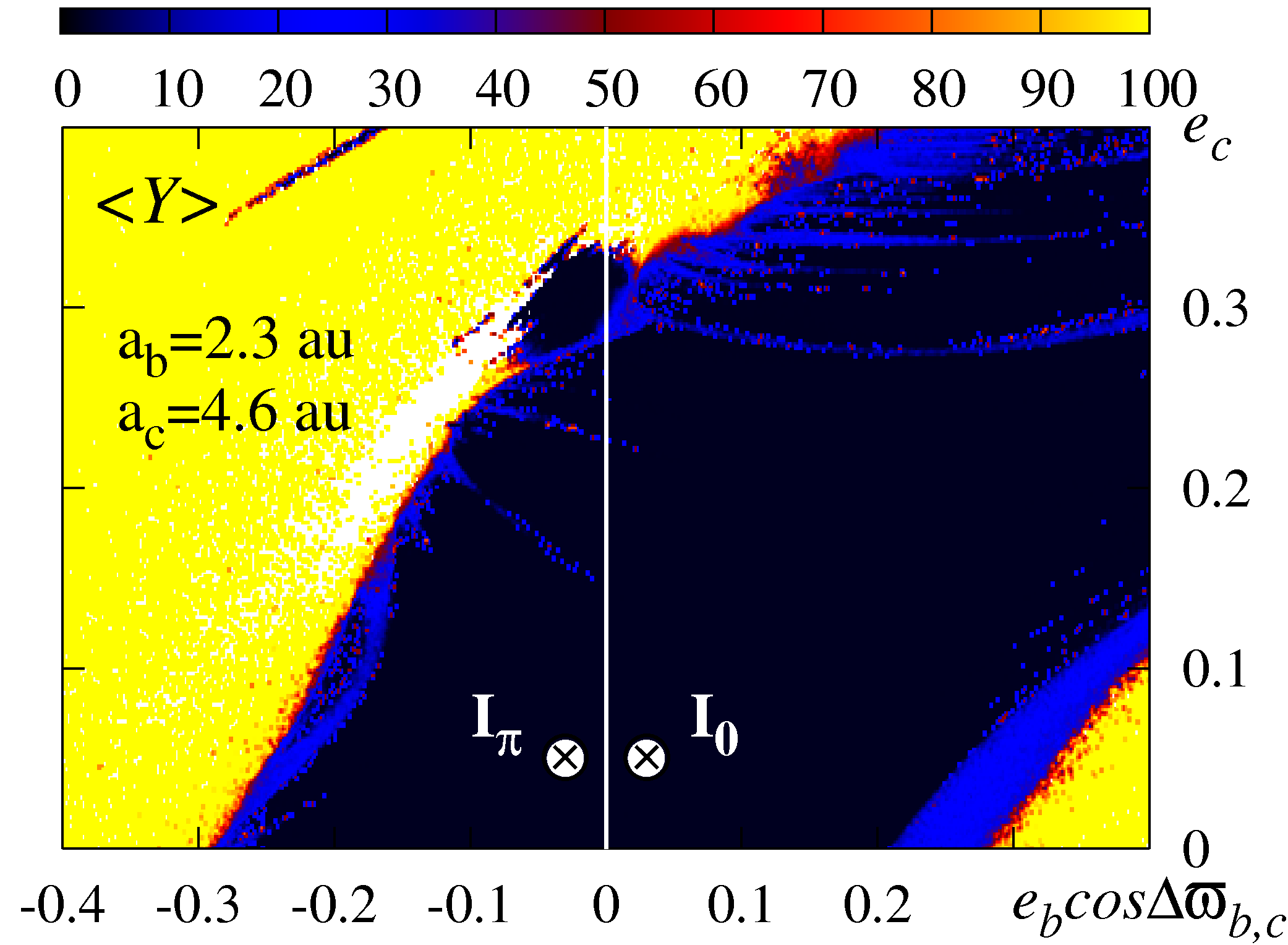}
              }
        \hbox{\includegraphics[  width=119mm]{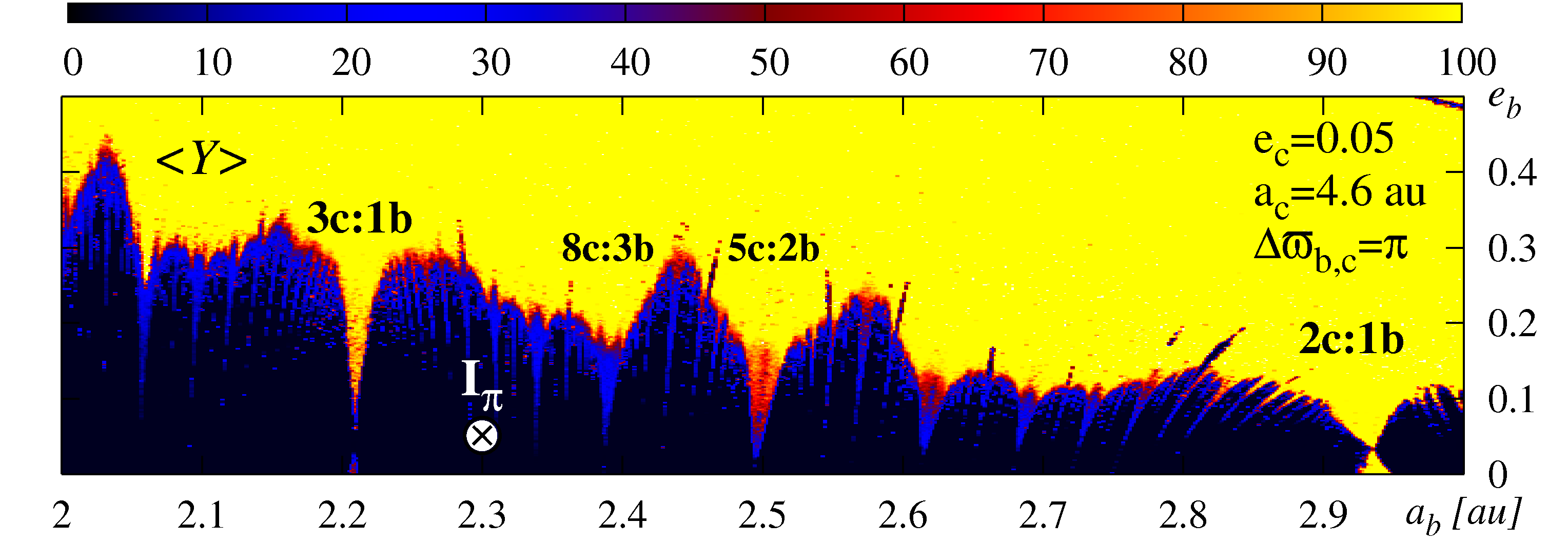}
              \includegraphics[  width=56mm]{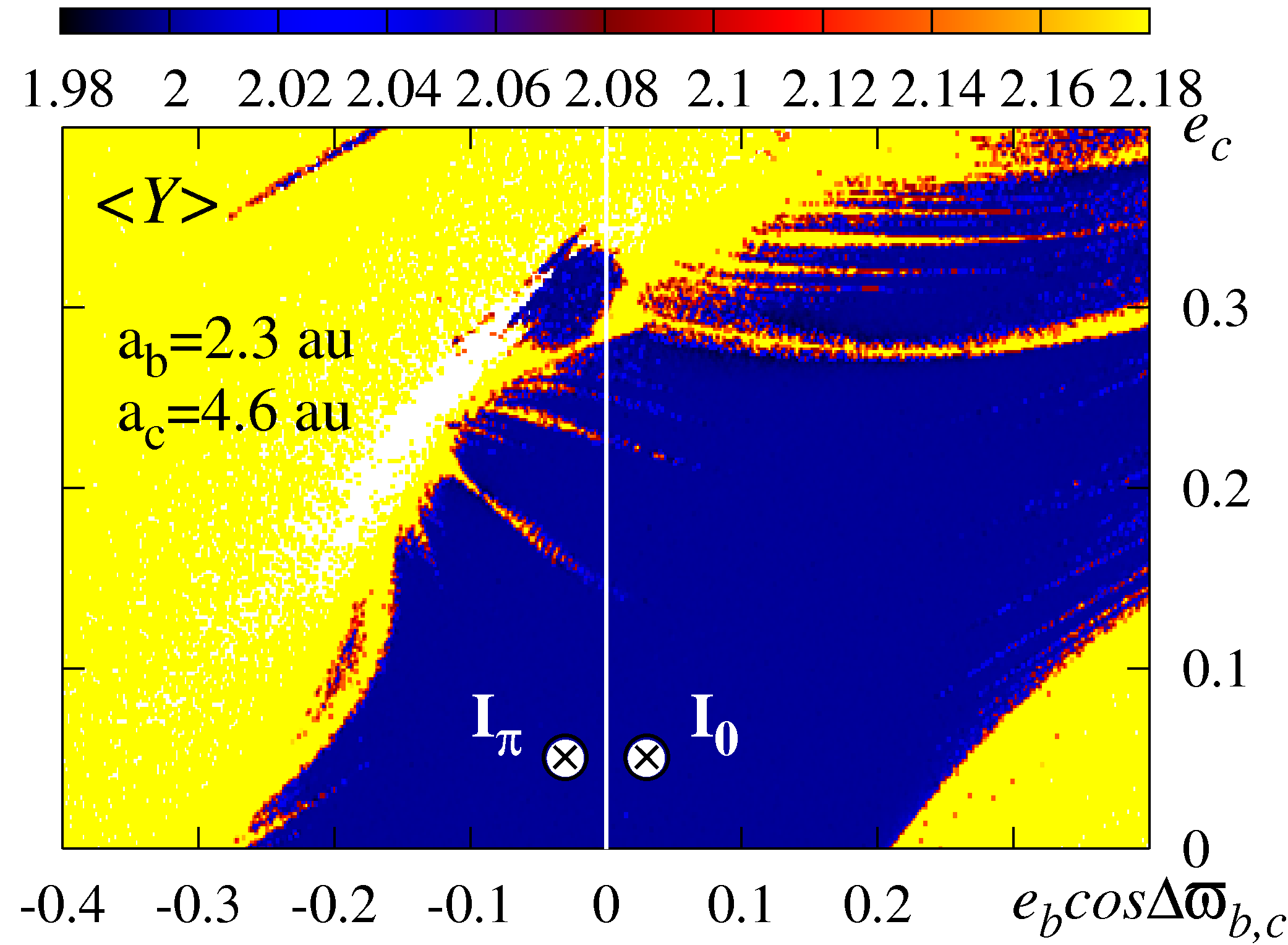}
              }
  }
}
\caption{
Dynamical maps of the \ogle{} system in terms of the MEGNO indicator, $\Ym$. The
type of orbits is color-coded; yellow means strongly unstable configurations of
jovian companions, black means quasi-periodic solutions. The integration time is
$\sim 10^5$ orbital periods of planet~c ($\sim 0.4$~Myr). The position of the
nominal system in the phase space is marked by crossed circle in each map.
Left-hand panels are for the ($a_{\idm{b}},e_{\idm{b}}$)-plane and
$\Delta\varpi_{\idm{b,c}}=0,\pi$, respectively. Panels in the right-hand column
are for the representative plane (see the text for details) and the MEGNO range
set to [0,100] (the top panel) and [1.98,2.18] (the bottom panel). A comparison
of these maps reveals some dynamical structures and transition zones between
strongly and mildly chaotic motions. Note a change of the color coding in the
bottom right-hand panel.
}
\label{fig:fig1}
\end{figure*}
In this work, we are heavily inspired by the ideas and results of
\cite{Malhotra2008}. To model the secular dynamics of the \ogle{} system,  they 
applied the classic linear Laplace-Lagrange (L-L) theory \citep[see,
e.g.,][]{Murray2000}, assuming non-resonant, quasi-circular orbits of the jovian
planets. Yet, as we shall demonstrate in Sect.~2, the orbital data combined with
the best-fit errors permit a variety of orbital states, including resonant or
highly eccentric configurations. In such circumstances, the L--L theory cannot
give proper estimates of the secular frequencies. In this work, we developed  a
relatively simple quasi-analytic algorithm that helps us to avoid the
limitations of the L--L theory.  We focus on the global dynamical structure of
the HZ, taking into account wide ranges of the orbital parameters  of jovian
companions permitted by the best-fit uncertainties. In particular, we analyse
the sensitive dependence of the secular motion in the HZ on unconstrained
apsidal angle $\Delta\varpi_{\idm{b,c}} \equiv
\varpi_{\idm{b}}-\varpi_{\idm{c}}$, where $\varpi_{\idm{b}}$ and
$\varpi_{\idm{c}}$ are the longitudes of periastra of planets~b and~c,
respectively. We also investigate the long-term stability of the \ogle{} system
by means of dynamical maps constructed with the fast indicator MEGNO, $\Ym$,
\citep{Cincotta2000,Cincotta2003}, and direct numerical integrations.
%
\section{Dynamics of the \ogle{} system}
\begin{figure*}
 \centerline{
    \hbox{\includegraphics [ width=56mm]{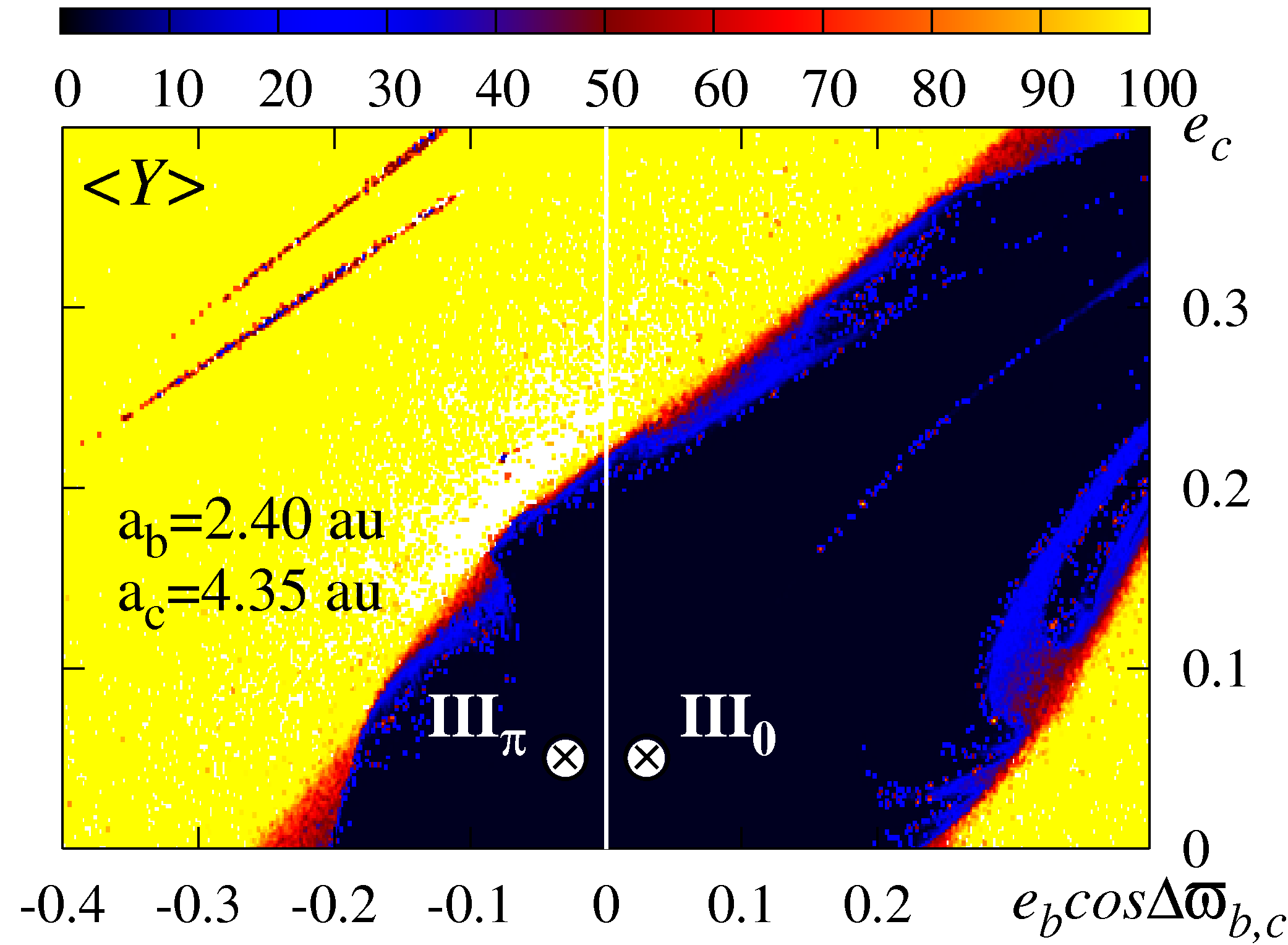}\hskip 3mm 
          \includegraphics [ width=56mm]{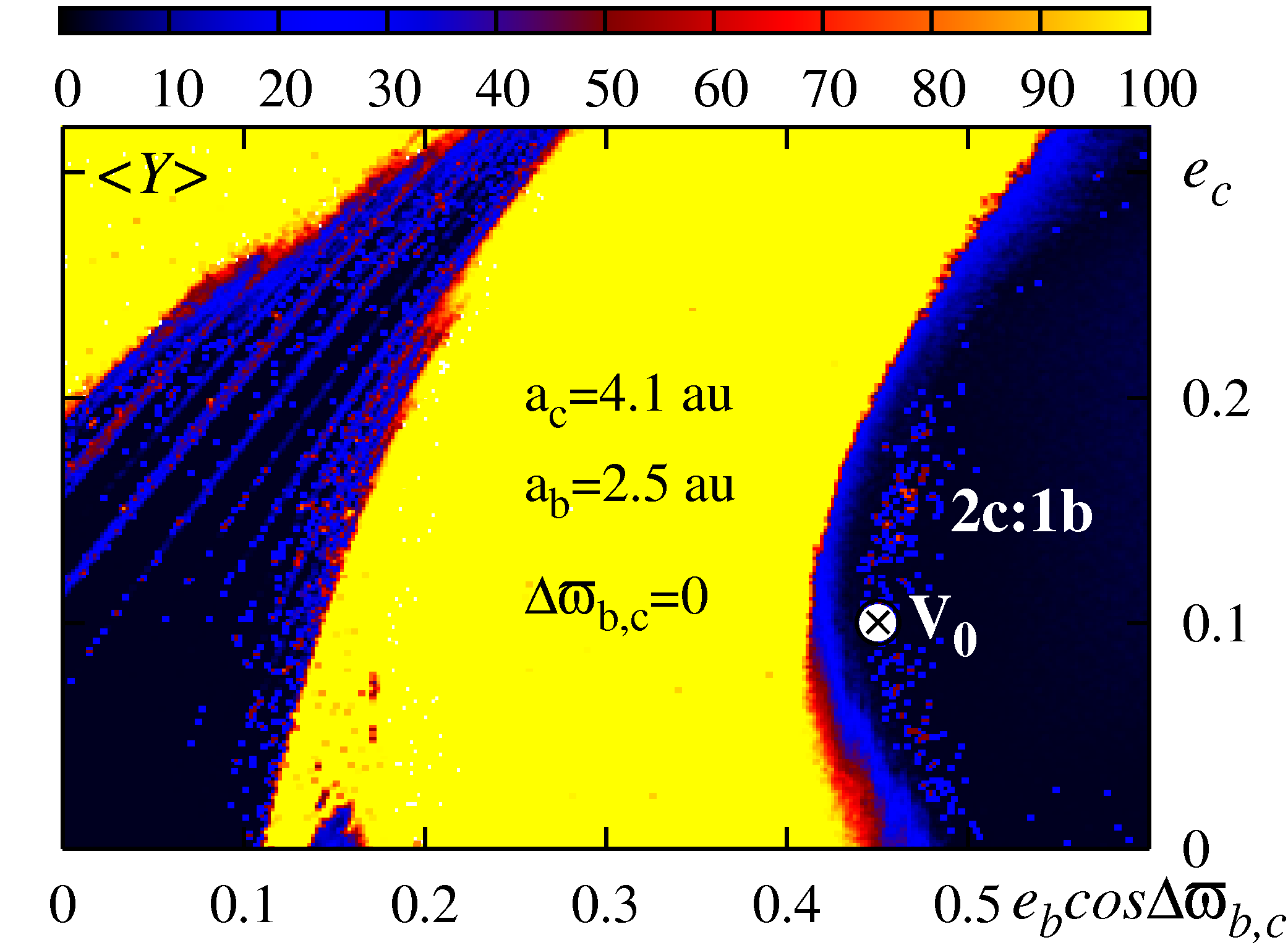}\hskip 3mm
          \includegraphics [ width=56mm]{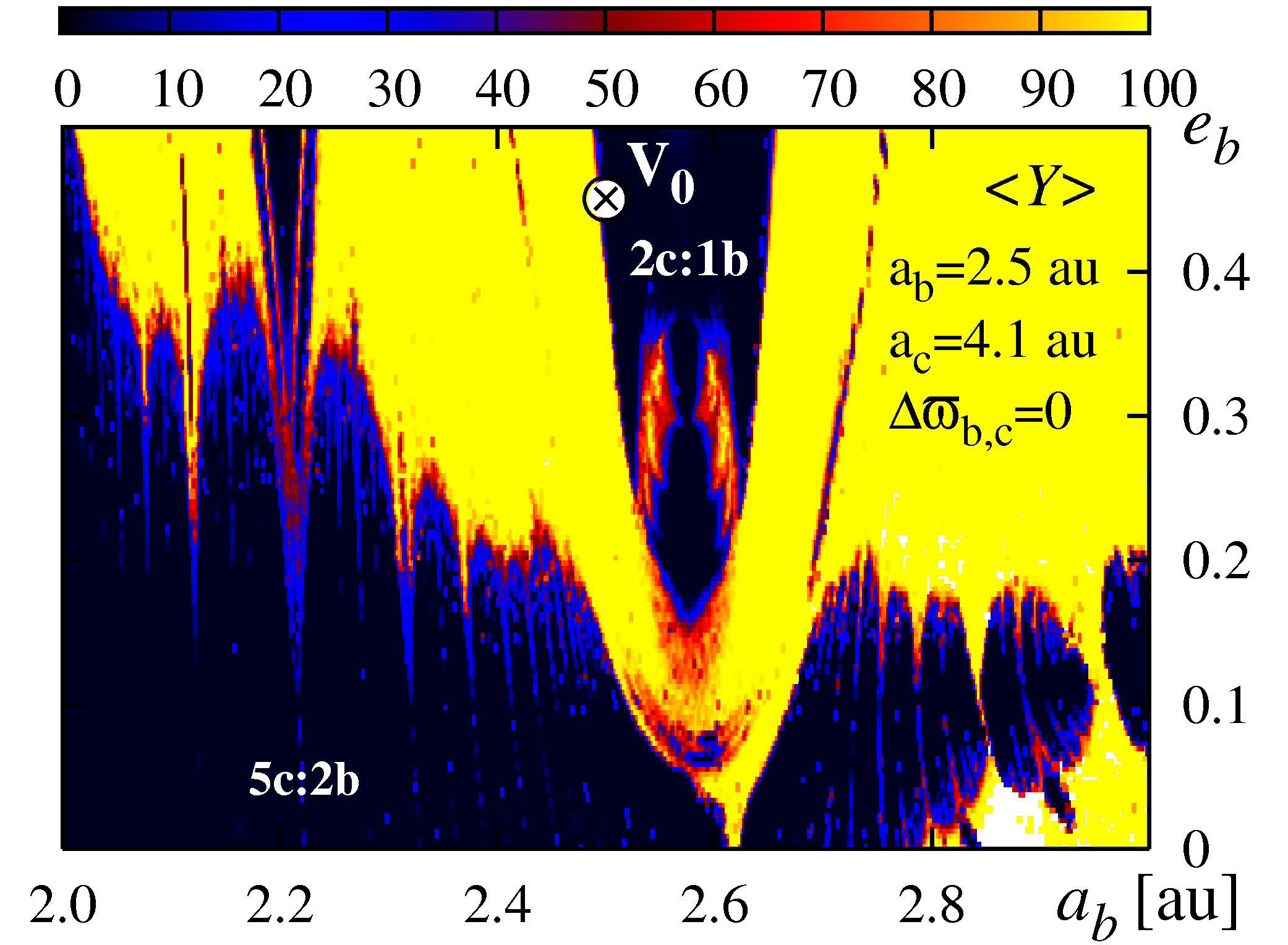} 
         }
}
 \caption{
Dynamical MEGNO maps for two putative configurations of the \ogle{} system. The
left-hand,  and the middle panels are for the ${\cal S}$-plane (see the text for
details), the right-hand panel is for the $(a_{\idm{b}}, e_{\idm{b}})$-plane.
Orbital elements of tested configurations are marked with crossed circle and
labeled.   The left hand panel is for model~$III$, the middle- and the
right-hand panels are for the 2c:1b MMR of Jupiters (model $V_0$).
}
\label{fig:fig2}
\end{figure*}

%
According to the discovery paper \citep{Gaudi2008}, the orbits of jovian
companions of \ogle{} are determined with significant uncertainties, i.e.,
$a_{\idm{b}} = 2.3 \pm 0.2$~au, $a_{\idm{c}} = 4.6 \pm 0.5$~au for the inner and
outer planet, respectively; $e_{\idm{b}}$ is unconstrained at all, and 
$e_{\idm{c}} = 0.11 \pm 0.07$.  To keep our work consistent with a
parameterization of the analytical theory in \citep{Migaszewski2008a},  we
interpret these orbital parameters as the canonical, geometric elements related
to the Poincar\'e coordinates   \citep[see,
e.g.,][]{Morbidelli2003,FerrazMello2006}. From the ``practical'' point of view,
these elements are not significantly different from the common, astrocentric
Keplerian elements.  The longitudes of pericenters ($\varpi_{\idm{b}},
\varpi_{\idm{c}}$), and  mean anomalies are undetermined. The inclination  of
\ogle{}c has been estimated from observations as $i \sim 59^{\circ}$, and also
masses of companions are constrained within $\sim 10\%$~error. It is reasonable
to assume that the whole system is coplanar. Still,  significant errors of the
semi-major axes leave  room for a few qualitatively different orbital
configurations which can be studied with the help of 2D dynamical maps. We
select and vary two Keplerian elements, and other orbital parameters are fixed. 
For each point of the parameter plane, representing an initial configuration of
the system, we compute the MEGNO signature \citep{Cincotta2000} with the
symplectic algorithm \citep{Gozdziewski2008a}.  A relatively short integration
time  required by MEGNO and good sensitivity of the indicator to chaotic
motions  make it possible to investigate large volumes of the phase-space with
high resolution, and to detect efficiently unstable configurations. For
instance, the dynamical maps in the $(a_{\idm{b}},e_{\idm{b}})$-plane,
illustrated in the left-hand panels of Fig.~\ref{fig:fig1}, have the resolution
of $640\times 400$ points and each point in these map represents a configuration
integrated over $\sim 0.4$~Myr. The orbital parameters of the nominal
configuration marked with crossed circle in the map are taken from
\citep{Malhotra2008}, and we choose $\Delta\varpi_{\idm{b,c}}$, following a
concept of the so called {\em representative plane} of initial conditions
\citep{Michtchenko2004}. That  $(e_{\idm{b}}
\cos{\Delta\varpi_{\idm{b,c}}},e_{\idm{c}})$-plane, ${\cal S}$ from hereafter,
is defined through fixed $\Delta\varpi_{\idm{b,c}} = 0$ (the right-hand
half-plane, for initially aligned orbits) and for $\Delta\varpi_{\idm{b,c}} =
\pi$ (orbits are anti-aligned). For these particular values of
$\Delta\varpi_{\idm{b,c}}$, the  eccentricities  in two-planet configuration
reach maximal or minimal values at the same time with an accord to the
conservation of the total angular momentum and the condition of
$\partial\Hsec/\partial\,\Delta\varpi_{\idm{b,c}}=0$ ($\Hsec$ is for the secular
Hamiltonian of the three-body problem). It can be shown \citep{Michtchenko2004}
that all phase trajectories of non-resonant two-planet system must intersect the
${\cal S}$-plane. Hence, the global dynamics of the system may be described 
conveniently in terms of initial conditions selected in the ${\cal S}$-plane.
The ${\cal S}$-plane is particularly useful for investigating equilibria of the
secular system. For a reference, other  examples of MEGNO maps in the ${\cal
S}$-plane are shown in the right-hand panels of Figs.~\ref{fig:fig1}
and~\ref{fig:fig2}. 

Clearly, the pair of Jupiters resides in dynamically active  region of the phase
space which is spanned by a few low-order MMRs. Allowing for a variation of
$a_{\idm{b}}$ within the $0.2$~au error, the \ogle{} system can be found in the 
neighborhood of the 3c:1b~MMR or 5c:2b~MMR. Because the error of $a_{\idm{c}}$
is $\sim 0.5$~au, even 2c:1b~MMR configurations are permitted. Simultaneously,
as the dynamical maps indicate, the border of stable zone strongly depends on
initial eccentricities.  Therefore, to determine the secular evolution of orbits
in the HZ, stable configurations of the giant planets must be chosen with care.
To illustrate that issue, we select a few representative configurations  (still,
consistent with the observational uncertainties) and compute their dynamical
maps. Orbital parameters of these models, in terms of parameter tuples $(
a_{\idm{b}} [\mbox{au}], a_{\idm{c}} [\mbox{au}], e_{\idm{b}}, e_{\idm{c}},
\Delta\varpi_{\idm{b,c}} )$, are given in Table~\ref{tab:tab1}: model $I_0$ for
the nominal configuration investigated in \citep{Malhotra2008},  with apsides
aligned and model $I_\pi$ with apsides anti-aligned; model $II_0$ with moderate
eccentricities, close to the best fit elements quoted in \citep{Gaudi2008};
models $III_{0}$ and $III_\pi$, for compact systems with moderate
eccentricities; models $IV_{0,\pi}$  for hierarchical configuration of the
Jupiters, and model $V_0$  involving  Jupiters in the 2c:1b~MMR. Dynamical maps
for models $III_{0,\pi}$ are shown in the left-hand panel of
Fig.~\ref{fig:fig2}, and for model $V_0$ in two remaining panels.

\begin{table}
\caption{
Orbital models of the \ogle{} system considered in this work. Masses of Jovian
planets are fixed to  $0.71$~m$_{\idm{J}}$ and $0.27$~m$_{\idm{J}}$  for
planets~b and~c, respectively, following Gaudi et al. (2008).  Mass of the
parent star is 0.5~$m_{\sun}$. The mean anomalies ${\cal M}_{\idm{b,c}}$
in these models are constrained by the mean longitude  $\lambda_{\idm{b,c}} =
\Delta\varpi_{\idm{b,c}} + {\cal M}_{\idm{b,c}}=0$.
}
\centering
\begin{tabular}{lccccc}
\hline
Model & $a_{\idm{b}}$~[au] & $a_{\idm{c}}$~[au]  
      & $e_{\idm{b}}$      & $e_{\idm{c}}$  & $\Delta\varpi_{\idm{b,c}}$ \\
\hline
$I_0$       & 2.30 & 4.60 & 0.03 & 0.05 & 0  \\
$I_{\pi}$   & 2.30 & 4.60 & 0.03 & 0.05 & $\pi$ \\
$II_{0}$    & 2.30 & 4.60 & 0.10 & 0.10 & 0 \\
$III_{0}$   & 2.40 & 4.35 & 0.10 & 0.10 & 0 \\ 
$III_{\pi}$ & 2.40 & 4.35 & 0.10 & 0.10 & $\pi$ \\ 
$IV_{0}$    & 2.10 & 5.10 & 0.10 & 0.10 & 0 \\
$IV_{\pi}$  & 2.10 & 5.10 & 0.10 & 0.10 & $\pi$ \\
$V_{0}$     & 2.50 & 4.10 & 0.45 & 0.10 & 0 \\
\hline
\end{tabular}
\label{tab:tab1}
\end{table}

%
\section{Semi-analytic model of the HZ}
\begin{figure*}
 \centerline{
    \hbox{\includegraphics [width=180mm]{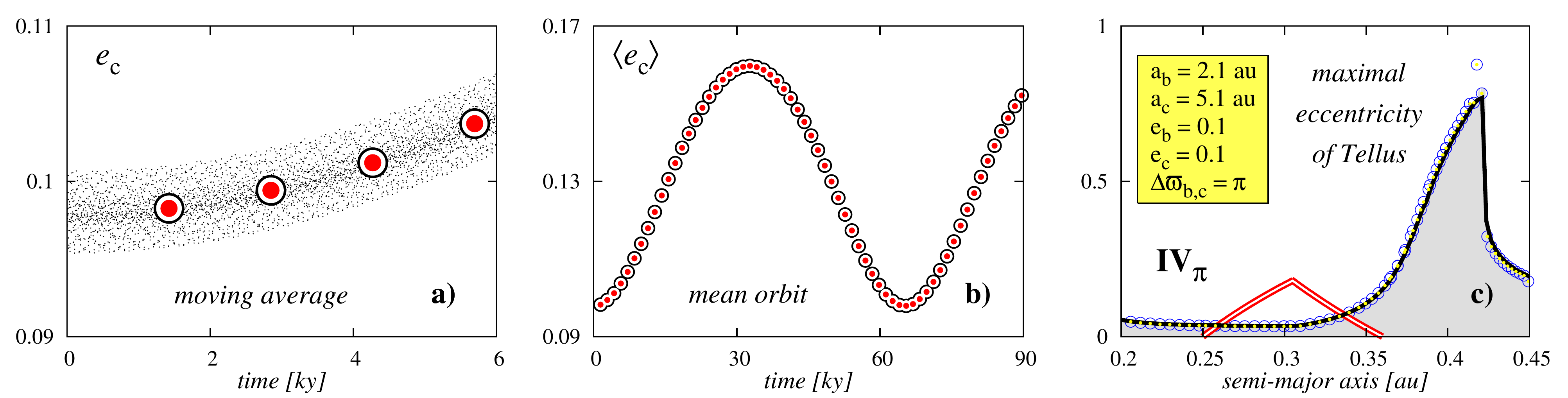}}
}
 \caption{
 The semi-analytic theory of Tellus explained  in graphic form. {\em Panel a)}.
 A part of $e_{\idm{c}}(t)$-curve (dots, the full three-body numerical
 integration), and the averaged eccentricity of the outer planet (red, filled
 circles). {\em Panel b)}. The mean eccentricity of the outer planet after
 averaging. {\em Panel c)}. A comparison of models of the $\nu_1$ secular
 resonance derived with the help of the semi-analytical theory (smooth, black 
 curve) and with the direct numerical integration of the restricted four-body
 problem (open circles).  Orbital parameters of the jovian sub-system are
 written in the box (model $IV_\pi$). Orbits with largest $\max e_0 \sim
 0.8$--$0.9$ are chaotic and that explains significant deviations of the
 numerical values from the semi-analytic solution. The red double curve marks
 geometric boundary of the HZ implied by Eq.~\ref{eq:eq1}. Initial conditions
 are: $\Delta\varpi_{0,c}=0$, $e_0=0.01$, and the  mean anomalies of Tellus
 ${\cal M}_0$ were chosen at random for the numerical integrations.
 }
\label{fig:fig3}
\end{figure*}

\begin{figure*}
 \centerline{
 \vbox{
    \hbox{\includegraphics [width=58mm]{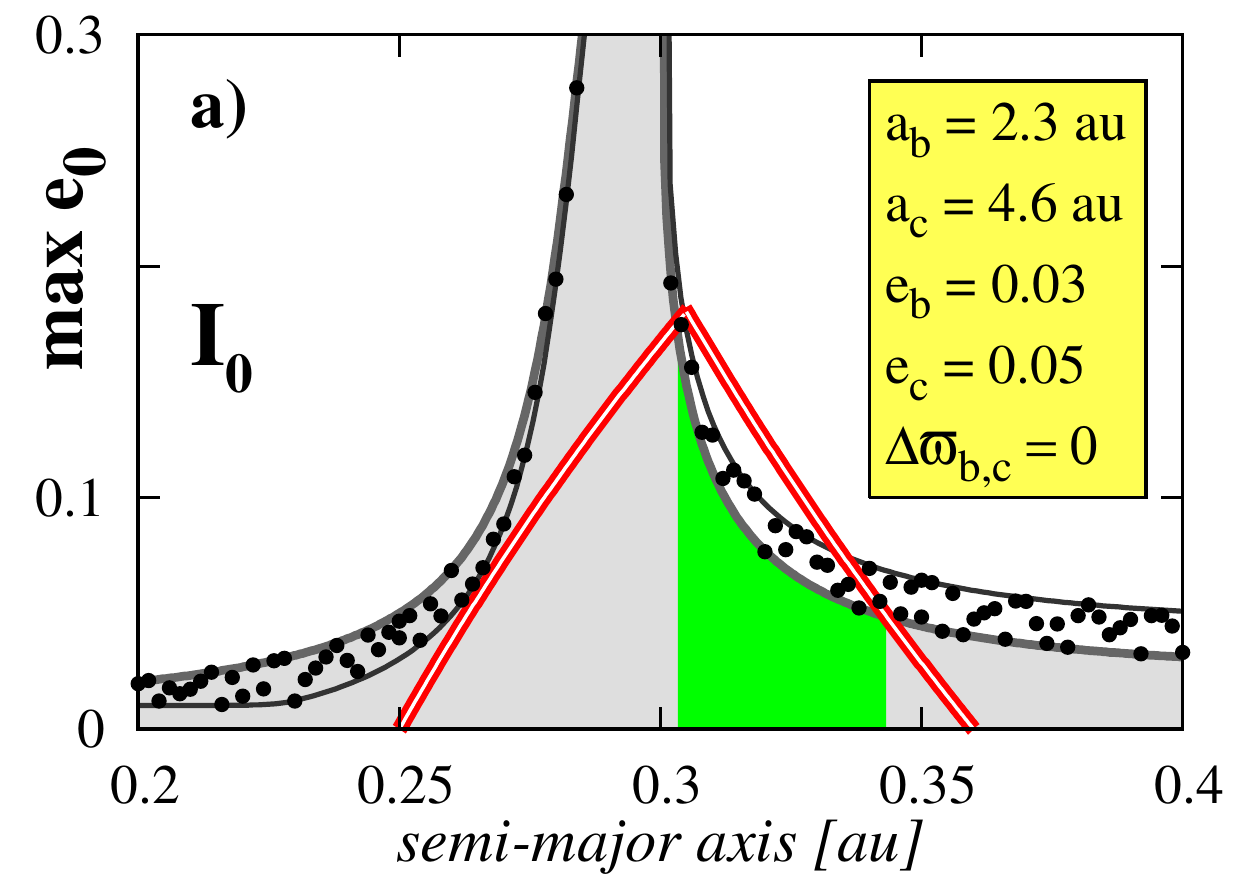} 
          \includegraphics [width=58mm]{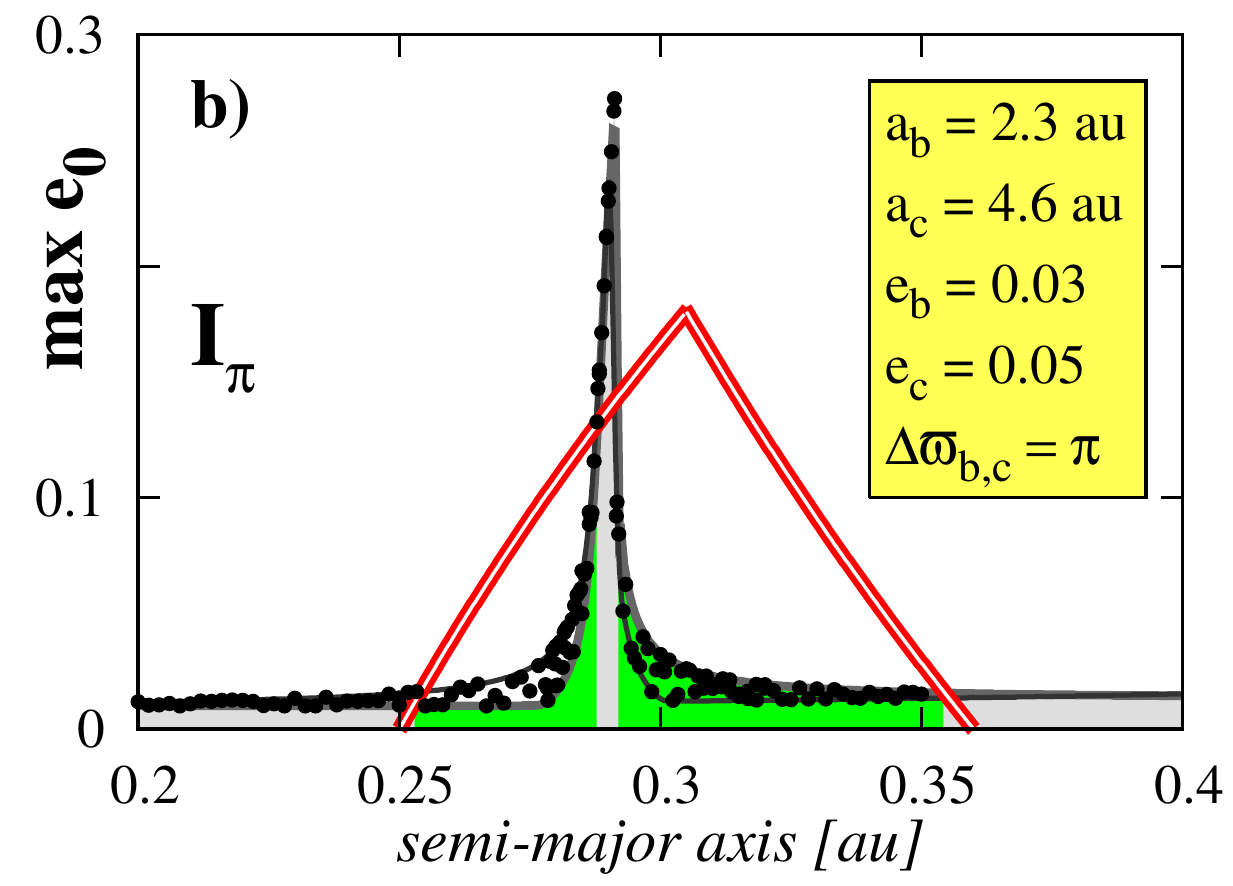}
          \includegraphics [width=58mm]{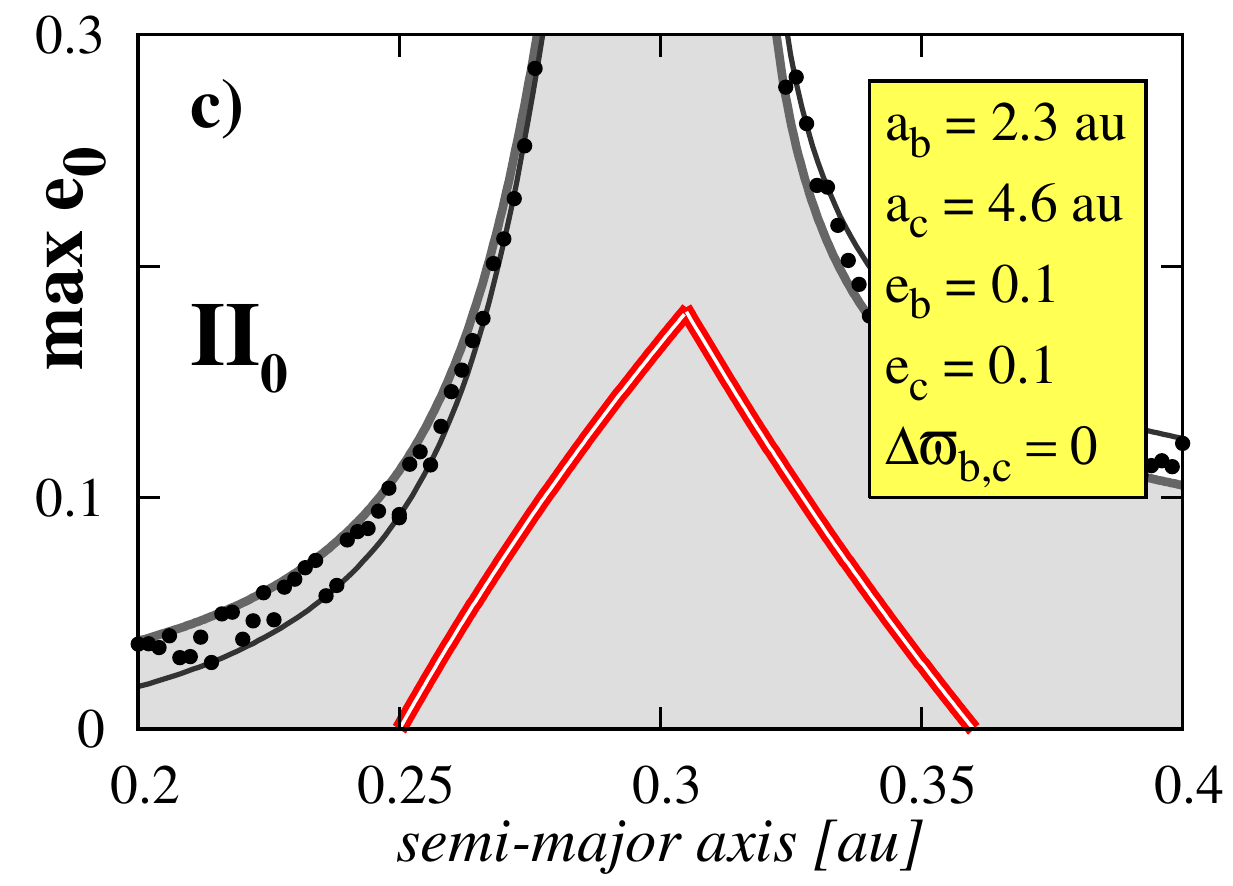}
         }
    \vskip 0mm
    \hbox{\includegraphics [width=58mm]{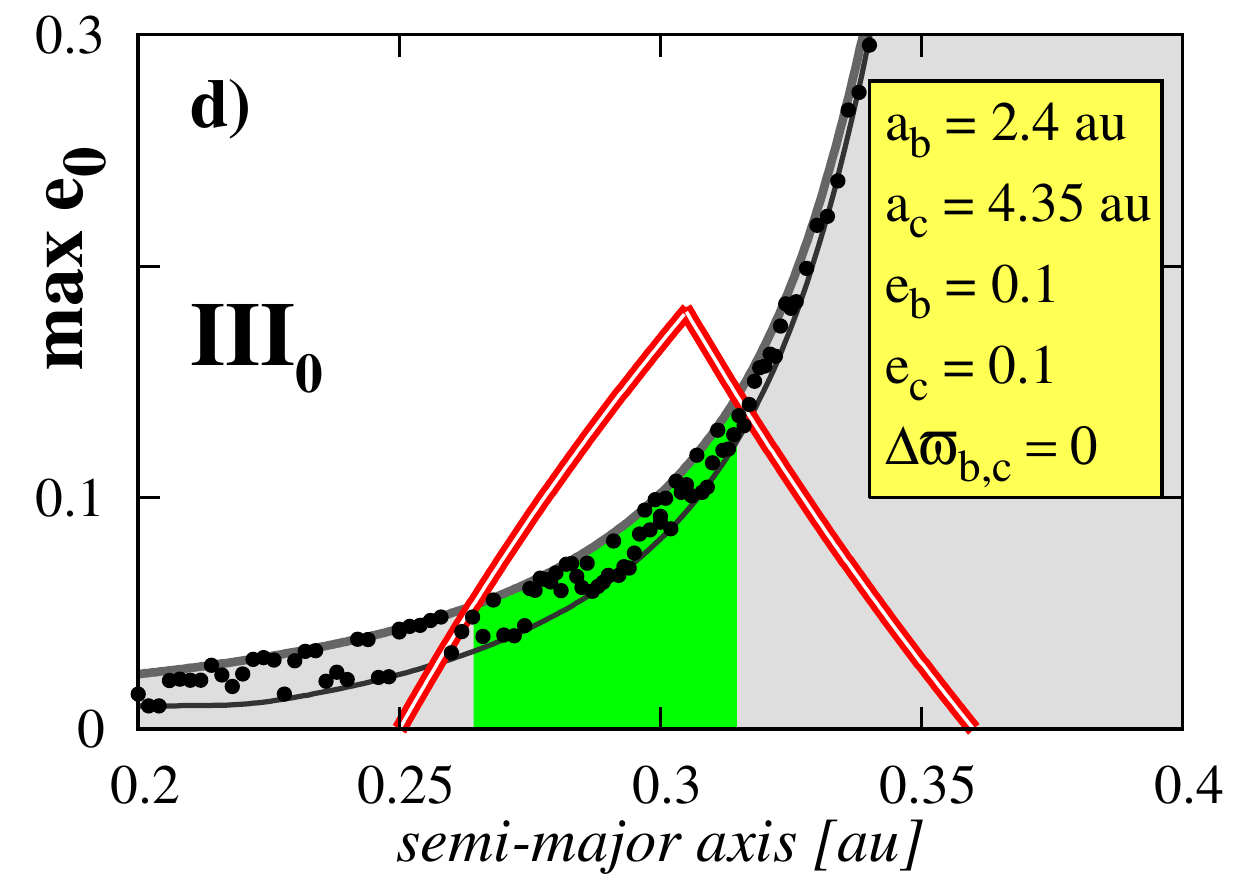} 
          \includegraphics [width=58mm]{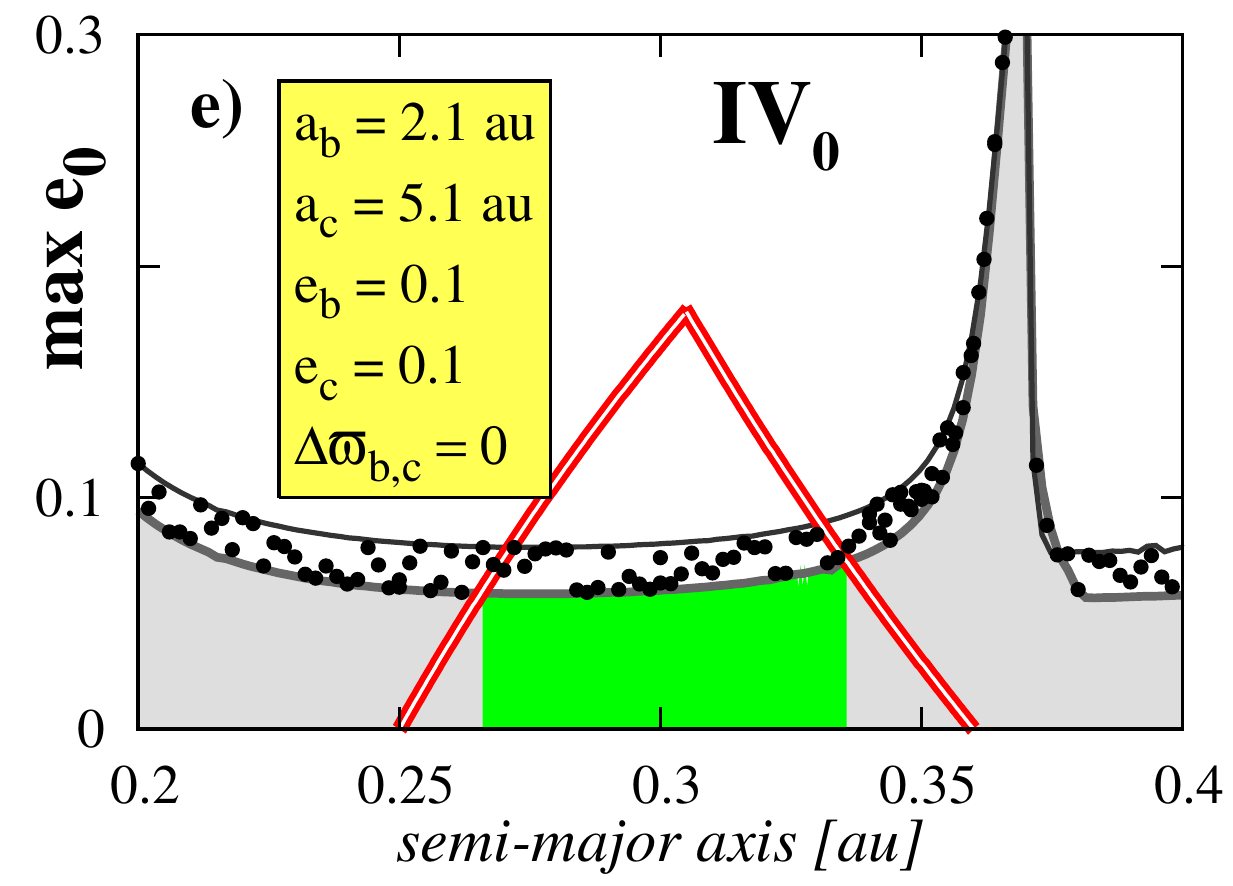} 
          \includegraphics [width=58mm]{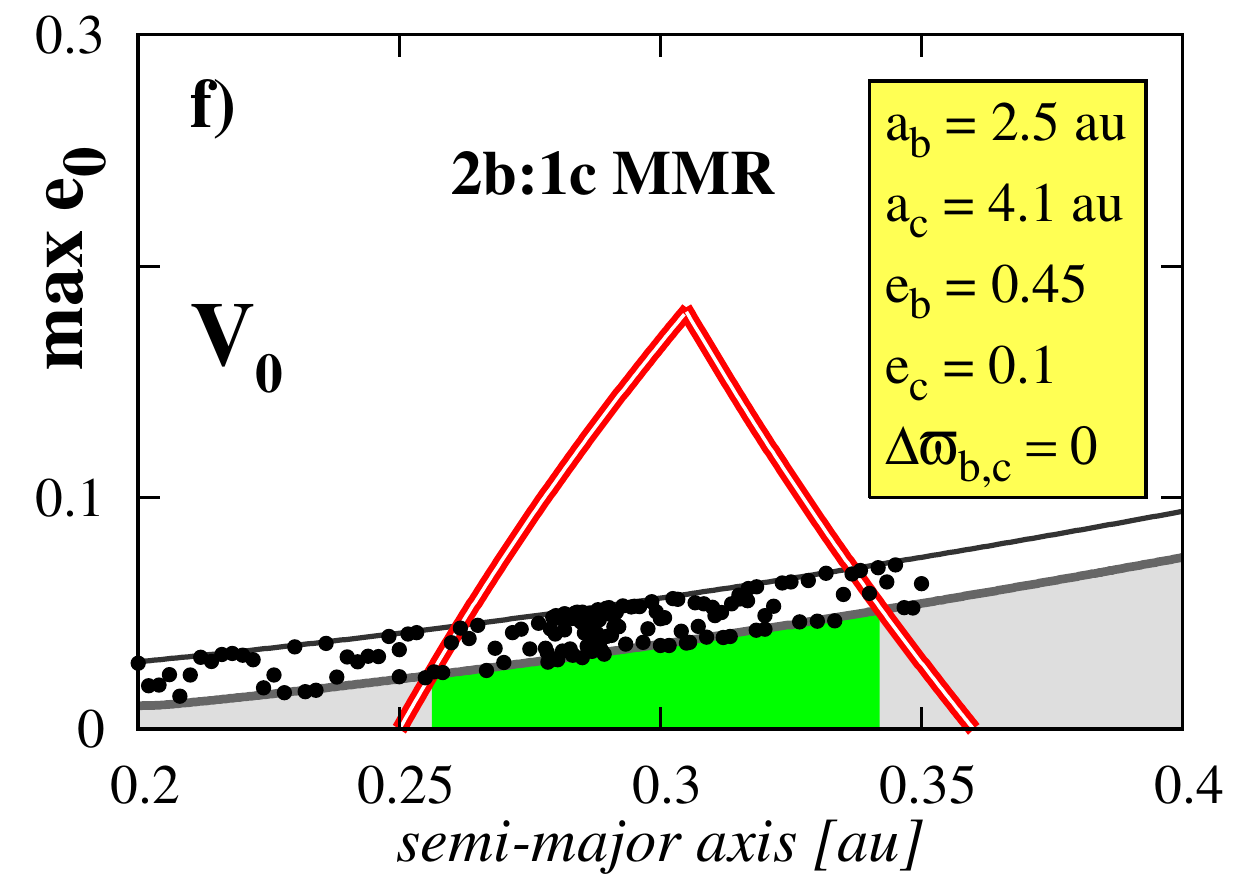}
         }
	 }
 }
 \caption{
The maximal eccentricity of Tellus as a function of initial semi-major axis
$a_0$. Initial eccentricity $e_0=0.01$. Mass parameters of the \ogle{} system
are: $m_0 = 0.5~\mbox{M}_{\odot}$,   $m_{\idm{b}} = 0.71~\mbox{m}_{\idm{J}}$,  
$m_{\idm{c}} = 0.27~\mbox{m}_{\idm{J}}$. Subsequent panels are for initial
orbital elements of the Jupiters expressed in terms of parameter tuples: $(
a_{\idm{b}} [\mbox{au}], a_{\idm{c}}[\mbox{au}], e_{\idm{b}},
e_{\idm{c}},\Delta\varpi_{\idm{b,c}} )$  labeled in the boxes (see also the text
and Table~\ref{tab:tab1} for details). Smooth grey/black curves are derived
through the semi-analytic theory (for initial $\Delta\varpi_{0,c}=0,\pi$,
respectively), filled circles are for the results of the direct numerical
integrations spanning $\sim 20$~Myrs for random $\Delta\varpi_{\idm{0,c}}$. 
Green areas mark ranges of $a_0$ providing stable orbits of Tellus  confined to
the annulus of HZ between 0.25~au and 0.36~au for $\Delta\varpi_{\idm{0,c}}=0$.
}
\label{fig:fig4}
\end{figure*}

%
The time-scales of orbital periods of giants and Tellus in the HZ are very
different.  Moreover, due to very long period of the $\nu_1$ resonance, of a few
Myrs ($\sim 10^8$ orbital periods of Tellus), the direct integrations of the
planetary equations of motion would require incredible amounts of CPU time.  
%
%
Therefore, we consider Tellus placed in the HZ and each planet in the
dynamically coupled pair of Jupiters as highly hierarchical system, because the
ratio of semi-major axes $a_0/a_{\idm{b,c}}\sim 1/10$ is  small. That makes it
possible to average out the short-term variations of the orbit of Tellus over
the mean longitude, and to approximate its {\em secular} evolution by means of a
24-order analytical theory in the semi-major axes ratio
\citep{Migaszewski2008a}.  Yet the theory cannot be directly applied to jovian
orbits, because their elements can be varied in wide ranges, and non-resonant as
well as resonant or near-resonant configurations are permitted.  The averaging
of the whole system, with strongly interacting giant planets cannot be done
uniformly \citep[see, for instance][]{Malhotra1989}.  Hence, we introduce two
simplifications to the problem. At first, we assume that Tellus does not affect
the orbital evolution of the giants, so we  consider {\em the restricted
problem}. Then the secular Hamiltonian (per mass unit) may be written as a sum
of two terms
$
{\cal H}_{\idm{sec}} = {\cal H}_{\idm{b}} + {\cal H}_{\idm{c}},
$
for interactions of Tellus with each Jupiter, separately, 
\[
{\cal H}_{i} = -\frac{k^2 m_i }{a_i} \left(1 + \sqrt{1-e_i^2}
\sum_{l=2}^{\infty}
{\left[\frac{\alpha_{0,i}}{1-e_i^2}\right]^l
\mathcal{R}^{(0,i)}_l(e_0,e_i,\Delta\varpi_{0,i})}\right),
\]
where $\alpha_{0,i}=a_0/a_i$ and $i=b,c$. The explicit formulae for  expansion
terms, ${\cal R}^{(0,i)}_l(e_0,e_i,\Delta\varpi_{0,i})$ can be found in
\citep{Migaszewski2008a}. The equations of motion implied by  ${\cal
H}_{\idm{sec}}$ refer to the {\em mean} orbital elements of Tellus {\em and} the
mean orbital elements of Jupiters as {\em parameters} of the model.  In the
realm of the restricted problem, we can derive the mean orbits of Jupiters (and
their perturbations) by  numerical averaging. The full, $N$-body equations of
motion of these planets are integrated numerically at least over of a few
$\nu_1$ periods ($\sim 10$ Myr). Then the temporal, osculating elements are
converted with appropriate constant time step-size to the mean elements through
the running average (see Figs.~\ref{fig:fig3}a,b). Because we intent to
integrate the {\em secular} equations of motion of Tellus, the time step of the
averaging ($\sim 10^3$~yr) should be a small part of the apsidal period of its
orbit. With this relatively long time step, as compared to the orbital periods,
we remove all fast, quasi-periodic variations of the osculating elements. The
mean orbits of jovian planets need to be computed only once (because we consider
the restricted problem), and then we can efficiently reconstruct secular orbits
of arbitrary number of  Tellus ``clones'', integrating  numerically the secular
equations of motion induced by ${\cal H}_{\idm{sec}}$.  The integration of a
single terrestrial orbit over the secular time scale (typically, a few Myrs) is
rapid, by {\em a few orders of magnitude} shorter than of the full system: the
CPU time of the analytical solutions counts in minutes, but to integrate $\sim
100$ orbits of Tellus during $\sim 20$~Myrs, the MERCURY code
\citep[][integrator RADAU]{Chambers1999}  spent {\em a week} on four 2GHz
AMD/Opteron CPU cores. A test of this approach is illustrated in 
Fig.~\ref{fig:fig3}c.  According with the classic L-L theory \citep{Murray2000},
the central peak of eccentricity $e_0$ appears when the particle's apsidal
frequency is close to the forcing planetary frequency. Hence that peak shows the
position of the resonance  in the parameter space (for instance, at the
$a_0$-axis). The position  of the eccentricity peak as well as the shape of
$\max e_0(a_0)$ graph are reproduced with great precision in the whole range of
$a_0$, in spite of extremely large eccentricity attained by terrestrial orbits
around $a_0 \sim 0.40$~au. Actually, after a few Myrs such orbits may become
strongly chaotic.
We did similar tests, choosing angle $\Delta\varpi_{0,c}$ and the mean anomaly  
${\cal M}_0$ at random. The results are illustrated in Fig.~\ref{fig:fig4}. At
this time, the orbits of   Jupiters are set as in models $I$--$V$ (see
Table~\ref{tab:tab1}). We compute the maximal $e_0$ of Tellus for varied initial
$a_0$ in the relevant range of $[0.2,0.4]$~au. We selected $\sim 100$ of clones
of Tellus, setting their initial $e_0=0.01$.   Next, we compared $\max e_0$
attained during $10$~Myrs, as computed with the help of the quasi-analytic
algorithm and  during 10--20~Myrs of the numerical integration of the restricted
four-body problem.  Figure~\ref{fig:fig4} is for one-dimensional plots of $\max
e_0$ as a function of initial $a_0$. Smooth curves obtained for
$\Delta\varpi_{\idm{0,c}} \equiv \varpi_0-\varpi_{\idm{c}}=0,\pi$ are for the
semi-analytic theory, filled circles are for the numerical integrations.
Elements of Jupiters are  labeled in boxes drawn in each respective panel and
with an appropriate Roman number referring to Table~\ref{tab:tab1}.  The results
for all tested configurations  perfectly coincide.  Filled points are found
strictly in the limits by analytical curves obtained for initial orbits of
Tellus aligned, and anti-aligned with  planet~b (or planet c). The position of
the $\nu_1$ resonance, the width of the eccentricity peak, and the maximal range
of $e_0$ are determined with great precision. These data are critical for
estimating a fraction of stable, quasi-circular orbits in the HZ: for a given
initial $a_0$ we compute the {\em secular} value of  $\max e_0$ and then we
compare that value with the limits (Eq.~\ref{eq:eq1}) implied by the orbital
annulus of the HZ. Its geometrical borders are marked with red solid curves in
Fig.~\ref{fig:fig4}; also the relevant range of semi-major axes, providing
secularly low-eccentric HZ-orbits for initial $\Delta\varpi_{0,c}=0$, are
marked  by projections of green areas under the $\max e_0$ curve onto the
$a_0$-axis. Panels of Fig.~\ref{fig:fig4}a,b are for the  configuration studied
by \cite{Malhotra2008}, i.e., $e_{\idm{b}} = 0.03, e_{\idm{c}} = 0.05$, and
$a_{\idm{b}} = 2.3$~au, $a_{\idm{c}} = 4.6$~au. Panel of Fig.~\ref{fig:fig4}a is
for aligned orbits (our model $I_0$), while Fig.~\ref{fig:fig4}b is for
anti-aligned orbits (our model $I_{\pi}$).   We can observe significant
differences between these  plots, in spite of that initial eccentricities and
semi-major axes are the same, and only initial relative orbital phase of jovian
planets (in fact, $\Delta\varpi_\idm{{b,c}}$ has changed).  That is quite a
different conclusion than could be derived in the framework of the L-L theory.
For $\Delta\varpi_{\idm{b,c}}=0$ a large part of the HZ is rendered unstable by
the secular resonance, nevertheless, for the anti-aligned configuration, most
particles would survive in the HZ, and $\max e_0(a_0)$ would be relatively
small. Moreover, the instability generated by the secular resonance (with the
centre at approximately 0.29 au)  is much stronger when initial eccentricities 
of Jupiters are $\sim 0.1$ (Fig.~\ref{fig:fig4}c). In such a case, $\max e_0$
can be as large as $0.8$--$0.9$ and all telluric orbits will be wandering far
out of the HZ. Similarly, for moderate $e_{\idm{b,c}}$ but different semi-major
axes of the jovian system (Fig.~\ref{fig:fig4}d,e,f), the position of the
secular resonance can be significantly shifted. Still, there will be some ranges
of $a_0$ permitting secularly stable motion of Tellus in the HZ (we recall that
these ranges are marked in green). A particularly interesting and quite
surprising result is illustrated in the last panel of Fig.~\ref{fig:fig4}f, that
refers to the 2c:1b~MMR of the jovian planets.  This MMR would protect Tellus
from the secular excitation of the eccentricity, through shifts of  the max
$e_0$ peak generated by $\nu_1$ far out of the HZ. Also this case confirms
excellent accuracy of the semi-analytic approach. 
As we can observe, even if the $\nu_1$ resonance disturbs the orbits of Tellus,
usually, there is a range of $a_0$, $\Delta a_0$, for which its orbit will
remain entirely in the HZ. According with the definition of the HZ, the maximum
range of $a_0$ is  $\max \Delta a_0 \sim 0.11$~au. To measure  a fraction of
HZ-stable orbits as a function of $a_0$, with respect to different configuration
of jovian planets, we define {\em the linear coefficient of habitability}: 
\begin{equation} 
\fHz = \frac{\Delta a_0}{\max \Delta a_0}, \quad \fHz \in [0,1],
\end{equation} 
where $\Delta a_0$ is the range of initial semi-major axis implying orbits
entirely confined to the HZ during the secular time-scale of a few Myrs. If the
whole HZ is rendered unstable by the $\nu_1$ resonance (or other perturbation)
then $\fHz= 0$, and if all orbits remain within the HZ annulus, we have $\fHz =
1$. For instance, for the nominal system in Fig.~\ref{fig:fig4}a, $\fHz \sim
0.4$.  The next configuration illustrated in Fig.~\ref{fig:fig4}b has $\fHz \sim
0.95$; in this case, the very narrow  $\nu_1$  resonance affects a small part of
the HZ. We can observe in Fig.~\ref{fig:fig4} that the position and  the width
of $\nu_1$ depends on assumed orbital  parameters of the jovian sub-system,
hence $\fHz \equiv \fHz(a_{\idm{b}}, a_{\idm{c}}, e_{\idm{b}}, e_{\idm{c}},
\Delta\varpi_{\idm{b,c}})$, also parameterized by  masses of large bodies $m_0,
m_{\idm{b}}, m_{\idm{c}}$. In general, $\fHz$ depends also on initial $e_0$ and
the relative phase of Tellus and a jovian planet, e.g.,
$\Delta\varpi_{\idm{0,c}}$.

\begin{figure*}
 \centerline{
 \vbox{
    \hbox{\includegraphics [width=72mm]{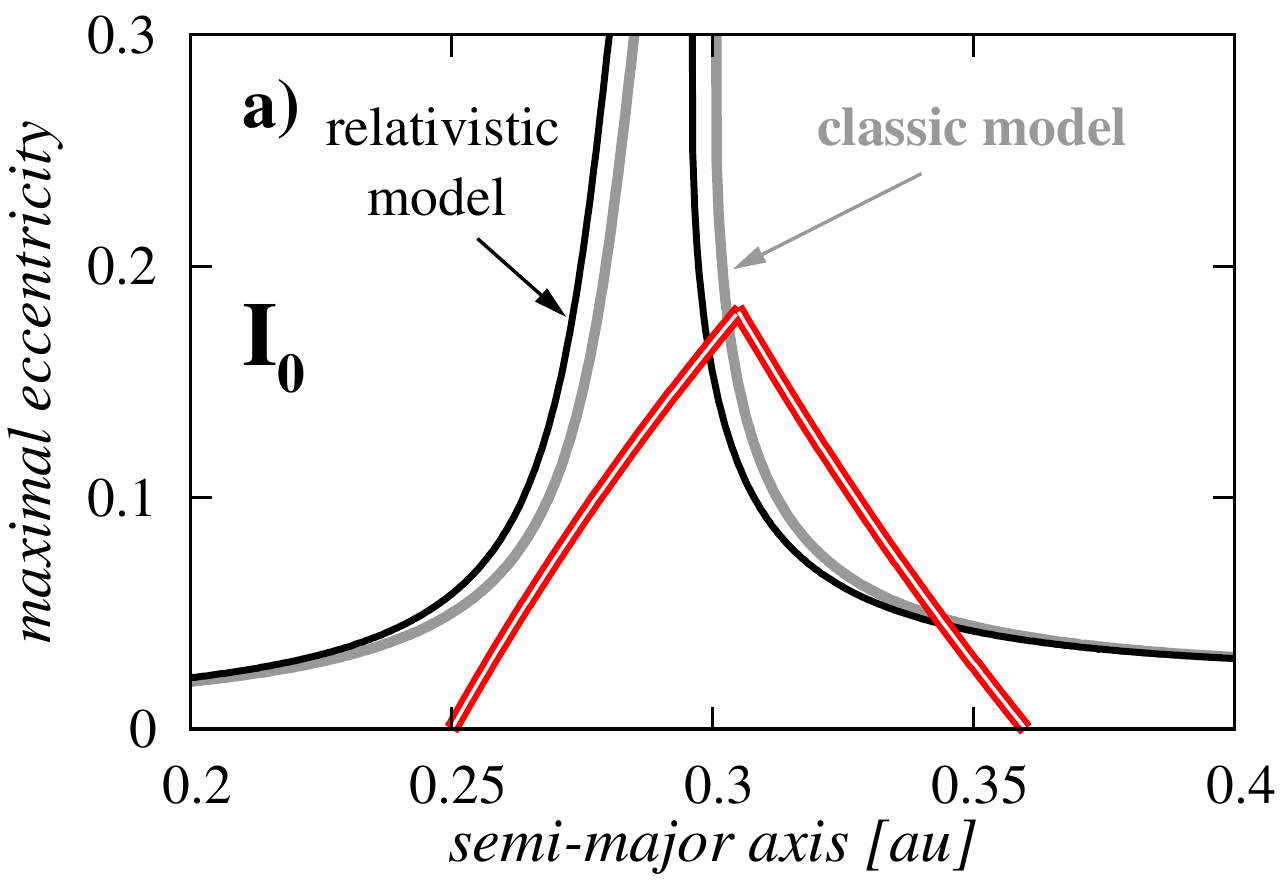} \hskip5mm
          \includegraphics [width=72mm]{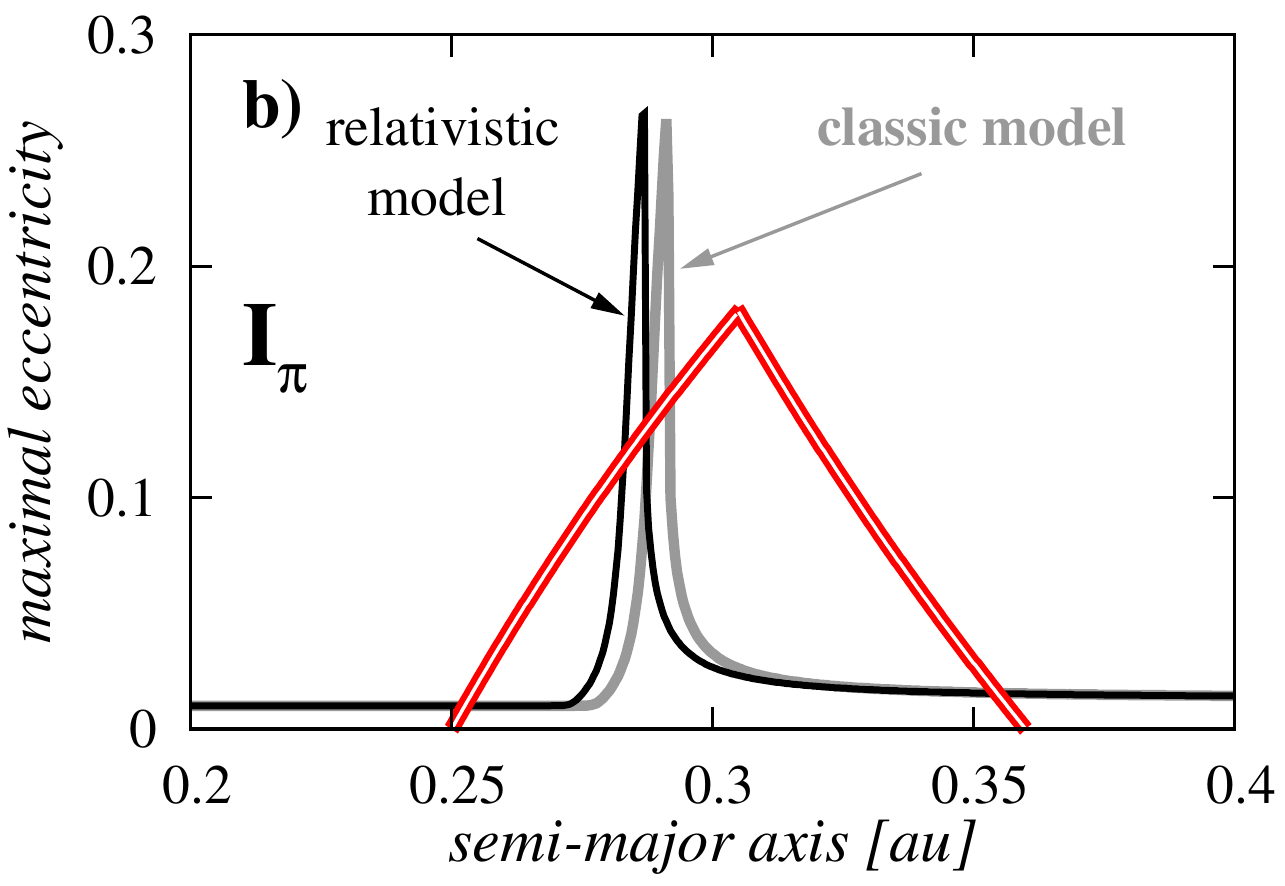}
         }
 }}
 \caption{
 A comparison of the maximal eccentricity of Tellus derived with the help of the
 semi-analytical theory, in the range of $a_0$ relevant for the HZ in model $I$.
 The initial eccentricity $e_0=0.01$ and $\Delta\varpi_{\idm{0,c}}=0$.  The
 left-hand panel is for $\Delta\varpi_{\idm{b,c}}=0$, the right-hand panel is
 for $\Delta\varpi_{\idm{b,c}}=\pi$. The  black curve is for the model with the
 general relativity corrections, the gray curve is for the classic, Newtonian
 point-mass model.
 }
\label{fig:fig5}
\end{figure*}

We also investigated the influence of the general relativity (GR) correction on
the secular dynamics of Tellus. In our recent paper \citep{Migaszewski2008b} we 
have shown that this apparently subtle correction to the Newtonian gravity may
affect the secular dynamics significantly, and they  can be particularly
important for small and close-in planetary companions. With the semi-analytic
model of the jovian orbits, we could repeat the calculations of $\max e_0$,
adding the GR correction to ${\cal H}_{\idm{sec}}$. The results derived with
these corrections ({\em the relativistic model}) and with the Newtonian
interactions only  ({\em the classic model}) are compared in
Fig.~\ref{fig:fig5}. For the nominal configuration, we observe only a small
shift of the eccentricity peak (of $\sim 0.01$~au). The maximal eccentricity is
not affected significantly.  Indeed, the GR corrections become important when
the rate of the apsidal motion, which it causes to change, is similar to the
effect of the Newtonian point-mass interactions. In the region of HZ, the GR
induced apsidal advance has the period of a few Myrs, while the Newtonian period
is $\sim 30,000$~yr only. Hence, we may conclude that the GR effects are not
really important for the  stable motion of Tellus in the HZ, and we do not
consider them anymore.

%
\section{Size of the HZ in the parametric space}
%
The one-dimensional parametric survey of the HZ with respect to $a_0$
(Fig.~\ref{fig:fig4}) can be generalized to the second dimension of  the initial
eccentricity $e_0$. Similarly to $\fHz$, we define {\em the planar coefficient
of habitability}: 
\begin{equation} 
\sHz \equiv \sHz(a_{\idm{b}}, a_{\idm{c}}, e_{\idm{b}}, e_{\idm{c}},
\Delta\varpi_{\idm{b,c}}) = \frac{\Sigma_0}{\max \Sigma_0}, \quad
 \quad \sHz \in [0,1],
\end{equation} 
where $\max \Sigma_0$ is the area of initial conditions in the given parameter
plane consistent with the  geometrical boundaries of the HZ, while $\Sigma_0$
measures the area of initial conditions in that plane, implying orbits confined
to the HZ during the secular time-scale. In contrary to $\fHz$, the planar
coefficient of habitability does not depend on $e_0$. Fixing
$\Delta\varpi_{\idm{0,c}}=0,\pi$, we follow again the concept of the
representative plane of initial conditions.

\begin{figure*}
 \centerline{
 \vbox{
    \hbox{\includegraphics [   width=58mm]{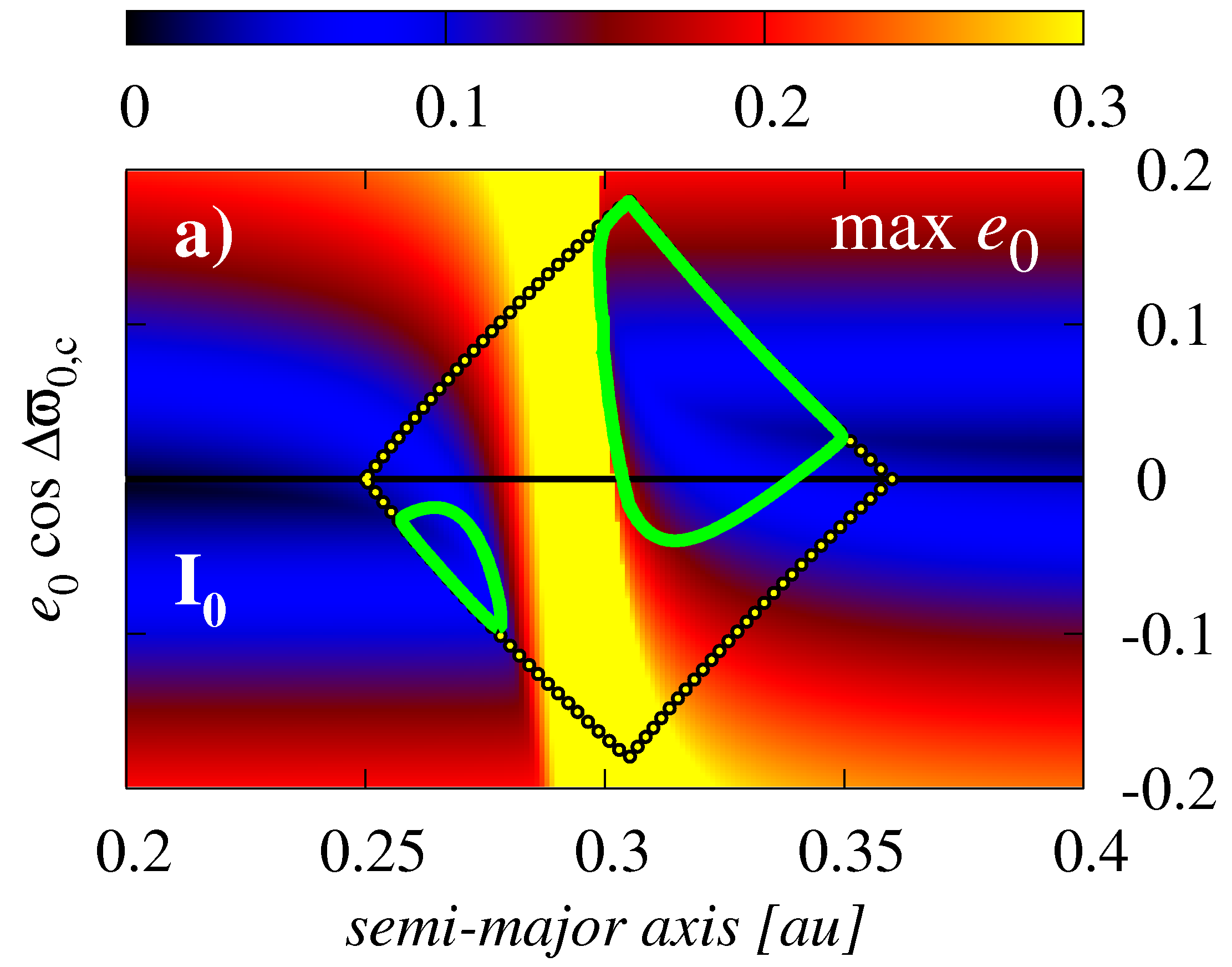} \hskip 1mm
          \includegraphics [   width=58mm]{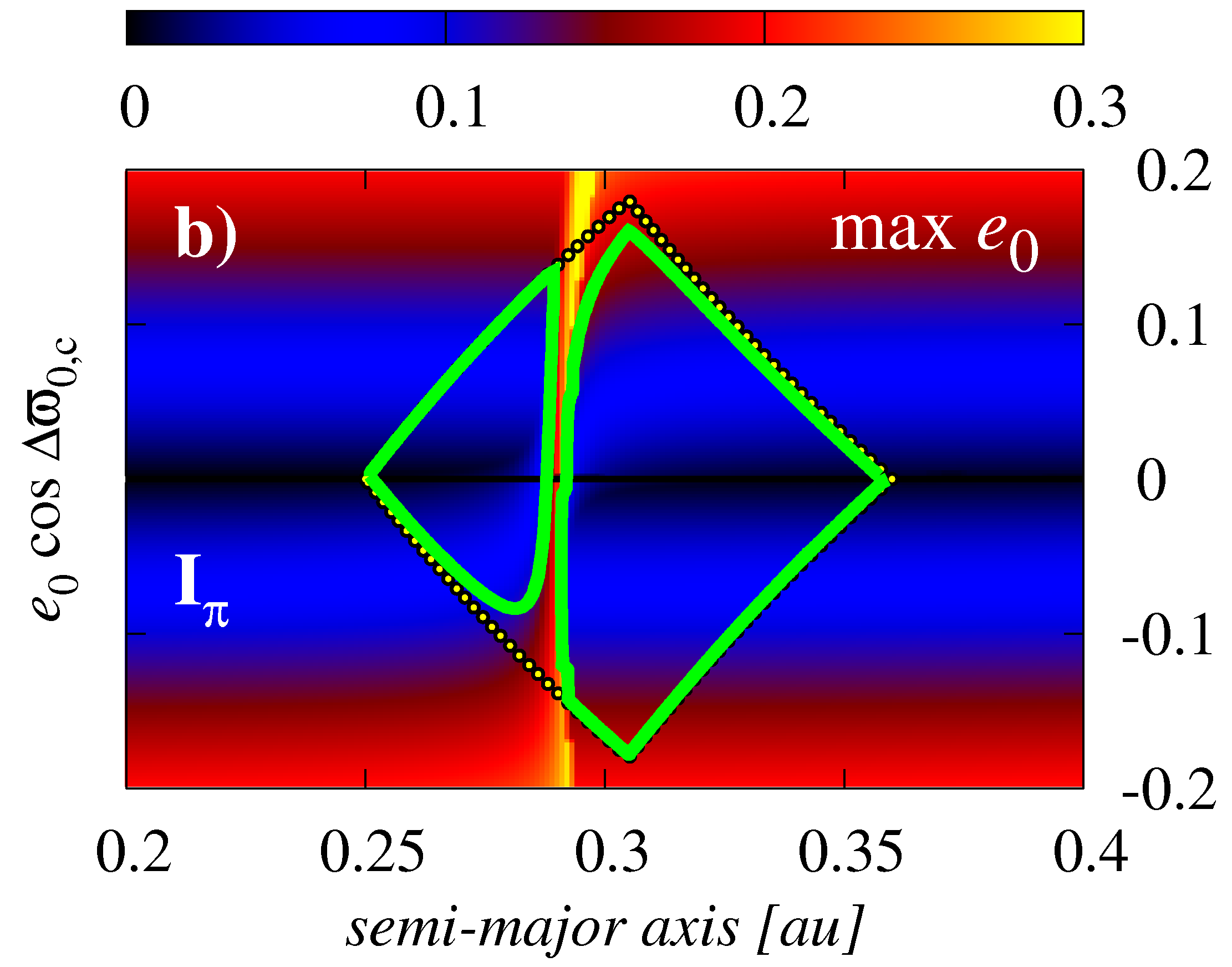} \hskip 1mm
          \includegraphics [   width=60mm]{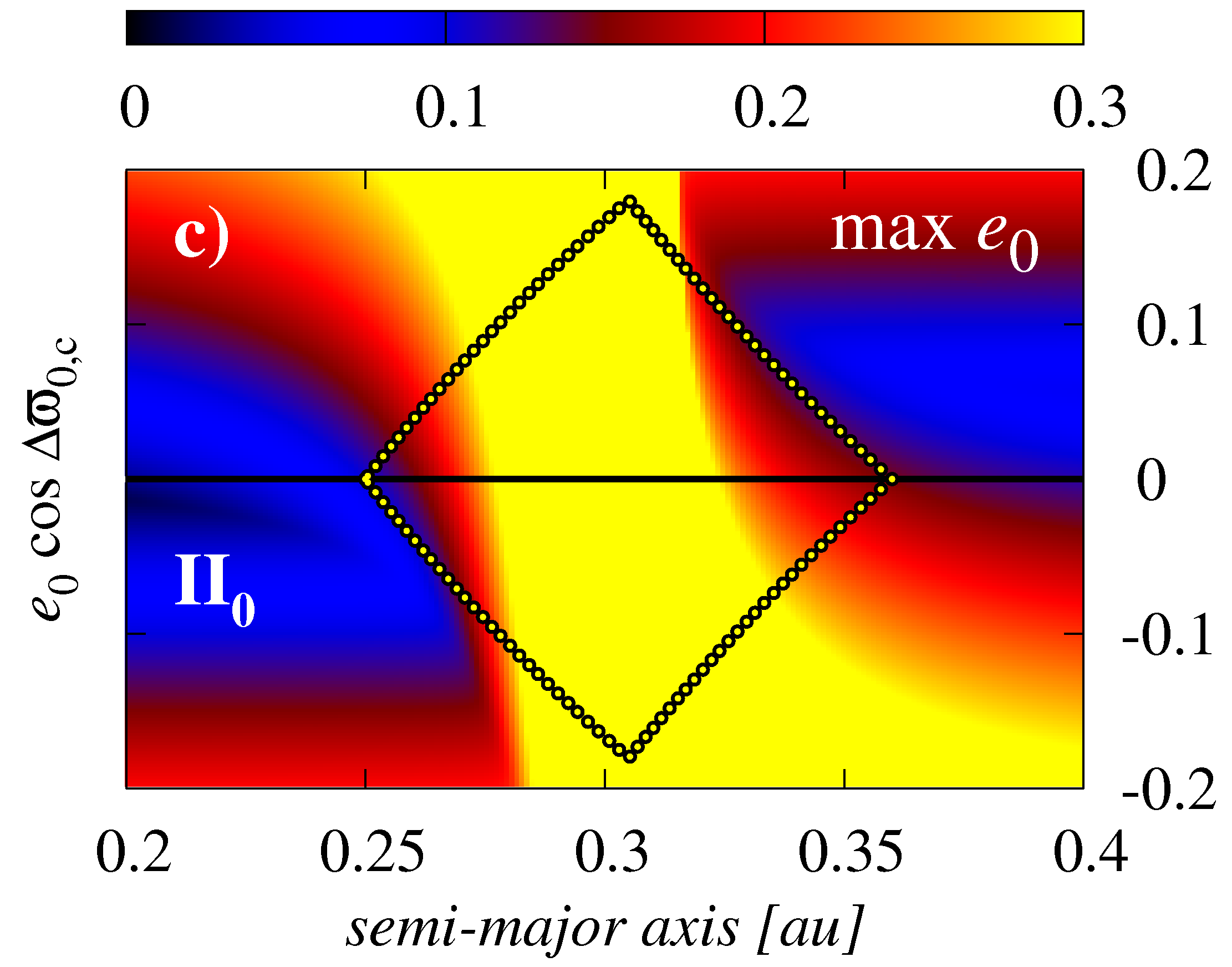}
         }
    \vskip 1mm
    \hbox{\includegraphics [   width=58mm]{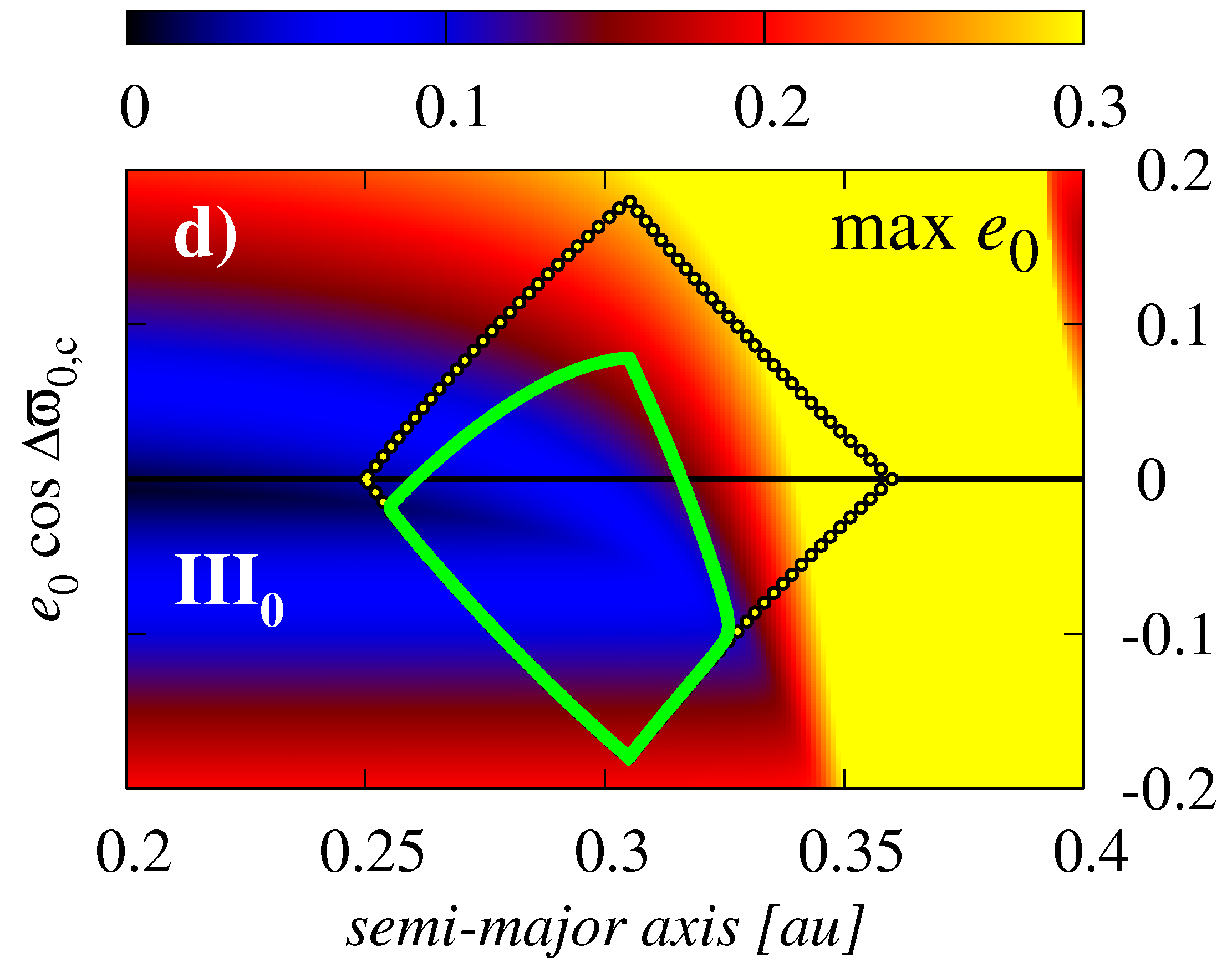} \hskip 1mm
          \includegraphics [   width=58mm]{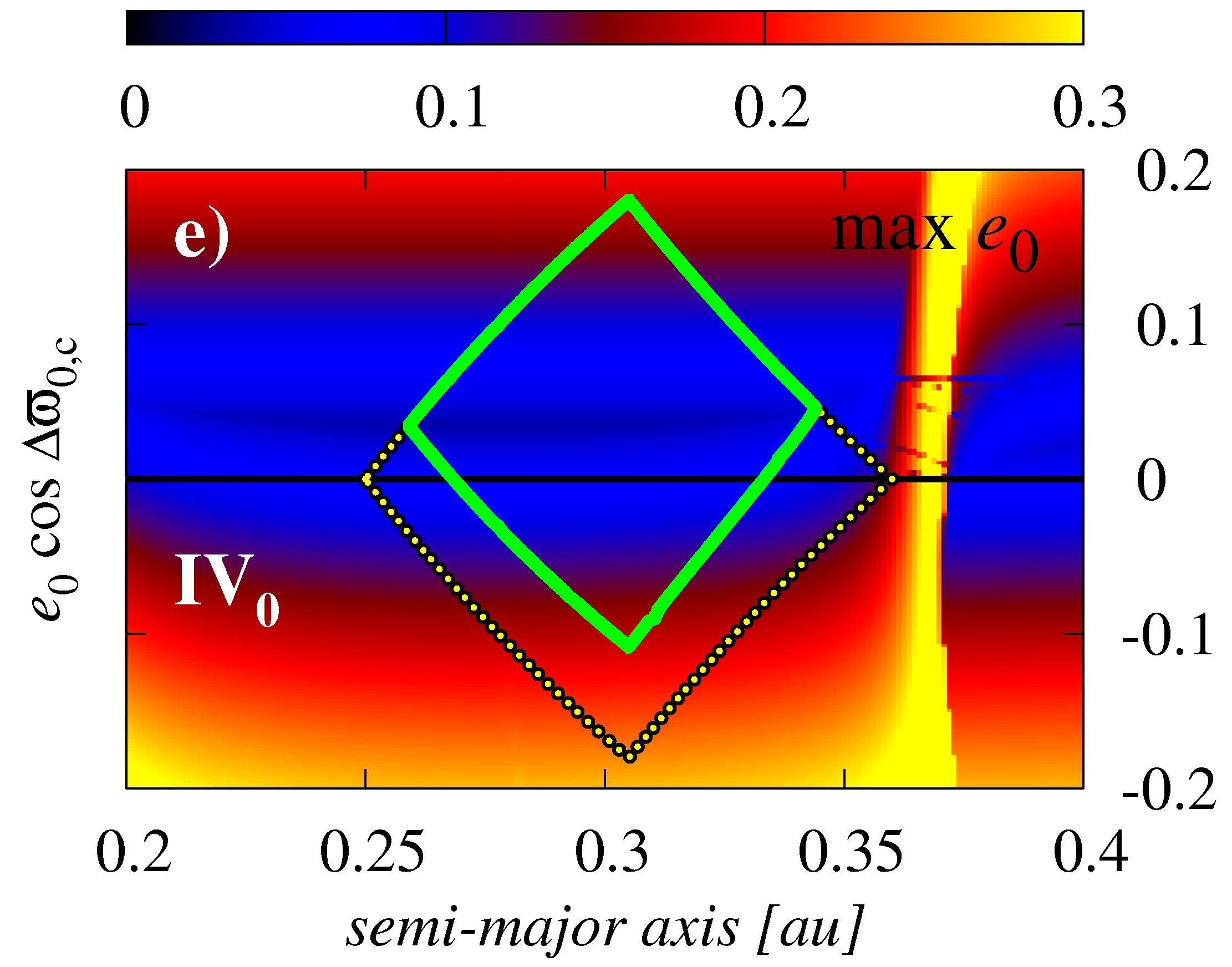} \hskip 1mm
          \includegraphics [   width=58mm]{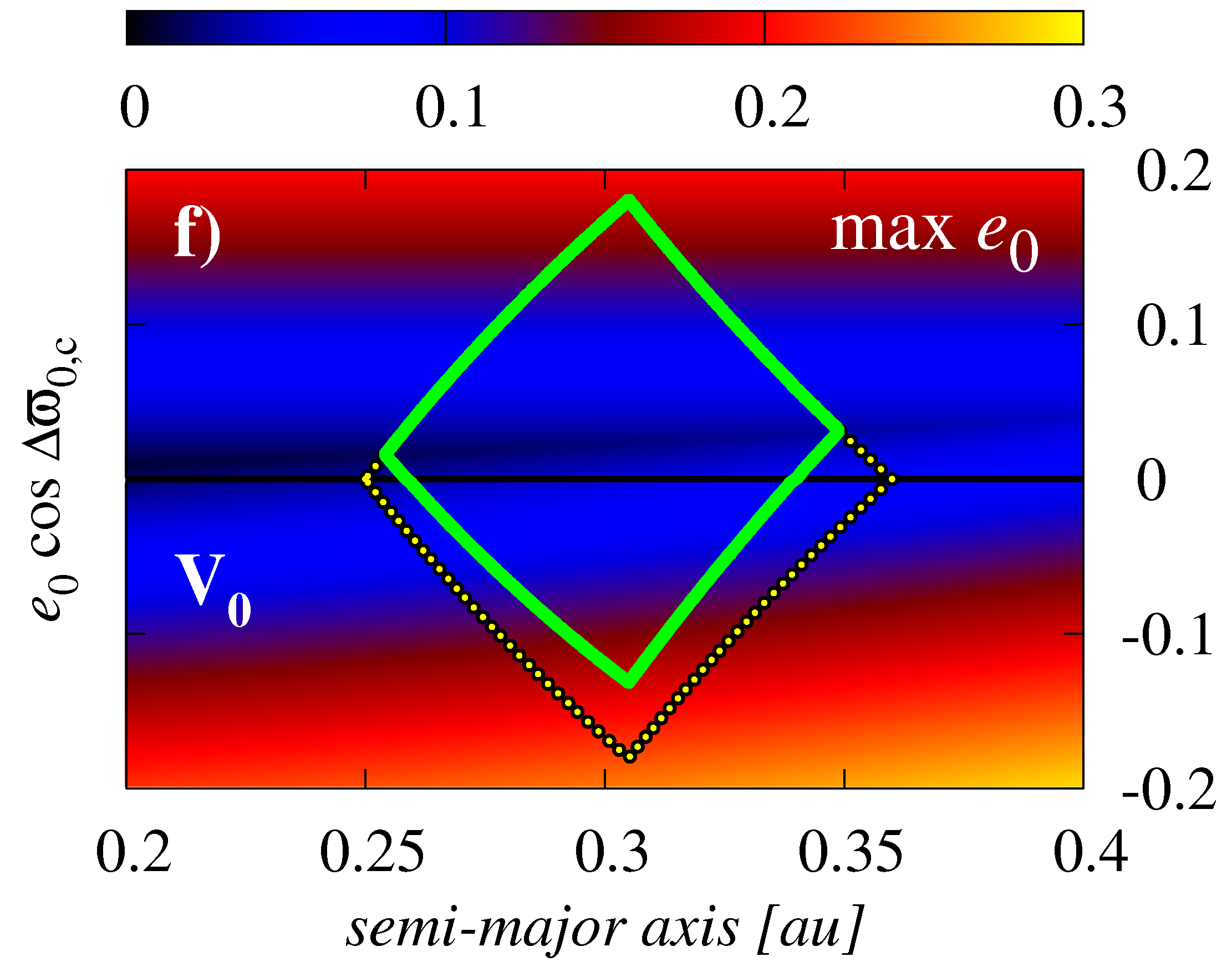}
         }
	 }
 }
 \caption{
 Color coded maximal eccentricity of Tellus in the   $(a_0,e_0
\cos\Delta\varpi_{\idm{0,c}})$-plane; $\Delta\varpi_{\idm{0,c}}=0$  in the
positive half-plane, and $\Delta\varpi_{\idm{0,c}}=\pi$ in the negative
half-plane. The thick, dotted curves mark geometric borders of the HZ with
accord to condition~(Eq. \ref{eq:eq1}, see the text for details). The green 
solid curves border initial conditions of Tellus implying orbits confined to the
HZ during at least $\sim 10$~Myrs. Subsequent panels are for the orbital
configurations of Jupiters given in Table~\ref{tab:tab1} and labeled
accordingly, see also Fig.~\ref{fig:fig4} (models $I$--$V$).
}
\label{fig:fig6}
\end{figure*}

The results for the same configurations of Jupiters as in Fig.~\ref{fig:fig4},
are illustrated in Fig.~\ref{fig:fig6}. Because the stable zone is bordered by
orbits of Tellus initially aligned or anti-aligned with the apsidal line of
planet~c (or, alternatively, with planet b), both these limiting cases can be
conveniently shown in one $(a_0, e_0 \cos\Delta\varpi_{\idm{0,c}})$-plane. The
negative values on the $y$-axis tell us  that $\Delta\varpi_{\idm{0,c}}=\pi$,
positive values mean that  $\Delta\varpi_{\idm{0,c}}=0$.  The borders of the
zone with HZ-stable orbits are marked with green solid curves. The geometric
boundary of $\max \Sigma_0$ is marked with thick dotted curves.
The stable zone has rather complex shape, and, depending on orbits of the jovian
sub-system, we can derive quite unexpected conclusions. For their apsides
aligned  (Fig.~{\ref{fig:fig6}a), there is still significant area of habitable
orbits for initial $e_0 \neq 0$. Curiously, if the jovian orbits are initially
anti-aligned, the HZ is almost wholly preserved  (Fig.~{\ref{fig:fig6}b). In the
next instance (panel~c),  the HZ is  rendered unstable, while in all remaining
cases (Fig.~{\ref{fig:fig6}d,e,f), there are HZ-stable orbits up to $e_0 \sim
0.15$.

Yet the tests illustrated in Figs.~\ref{fig:fig4} and \ref{fig:fig6} still
provide  limited information on the HZ because we analyzed only isolated
configurations of Jupiters. To obtain better insight into the dependence of the
HZ on different orbital setups of Jupiters, we extend the survey to computation
of  $\sHz$ in two-dimensional parameter planes of the jovian sub-system. 
%
\subsection{The HZ in the representative plane}
%
The features of the ${\cal S}$-plane make it particularly convenient to analyse
$\sHz$ for fixed semi-major axes of the jovian sub-system and its dependence on
the eccentricities. The results are illustrated in the left-hand panel of 
Fig.~\ref{fig:fig7}, which shows  color-coded values of $\sHz$ in the ${\cal
S}$-plane constructed for the nominal system ($a_{\idm{b}}=2.3$~au,
$a_{\idm{c}}=4.6$~au, our model $I$). It can be regarded as  non-resonant or
near-resonant (see Fig.~\ref{fig:fig1}). In such a case, in the regime of
moderate eccentricities, the apsidal angle $\Delta\varpi_{\idm{b,c}}$ circulates
or librates around $0$ or $\pi$ (these regions are dotted in
Fig.~\ref{fig:fig7}).  For a reference, the red, thick curves mark approximate
positions of stable stationary solutions of the  non-resonant secular model of
the jovian sub-system, calculated with the help of  a high-order analytic theory
\citep{Migaszewski2008a}.  The curve in the positive half-plane of ${\cal S}$ is
for the so called mode~I equilibria, characterized by librations of
$\Delta\varpi_{\idm{b,c}}$ around $0$ for close phase trajectories; mode~II
curve (in the negative half-plane of ${\cal S}$) is surrounded by orbits with
secular angle $\Delta\varpi_{\idm{b,c}}$ librating around $\pi$ [see
\citep{Michtchenko2004} for details]. The right-hand panel of
Fig.~\ref{fig:fig7} is for the phase diagram constructed for a fixed angular
momentum in the $(e_{\idm{b}} \cos\Delta\varpi_{\idm{b,c}}, e_{\idm{b}}
\sin\Delta\varpi_{\idm{b,c}})$-plane, illustrating  the equilibria and
neighboring phase trajectories, including the nominal configuration marked with
open circles. White regions in the left-hand panel of Fig.~\ref{fig:fig7} 
indicate eccentricities and $\Delta\varpi_{\idm{b,c}}$ for which the HZ is
strongly affected by the $\nu_1$ resonance, hence $\sHz=0$. Black regions are
for $\sHz \sim 1$. Before integrating secular orbits of Tellus, we must average
out orbits of  primaries, hence we have also ``an occasion'' to eliminate
unstable configurations (disrupted during $\sim 10$~Myrs, $\Ym>10$). They are
marked with gray crosses.  Nominal eccentricities of the Jupiters are marked
with filled circles (labeled with $I_0$ for $\Delta\varpi_{\idm{b,c}}=0$, and
$I_\pi$ for $\Delta\varpi_{\idm{b,c}}=\pi$). For the right-hand half-plane,
$\sHz \sim 0.5$, and for the left-hand half-plane of ${\cal S}$, $\sHz \sim
0.95$. Of course,  it must be in accord with the results of two-dimensional
surveys (see Fig.~\ref{fig:fig6}) performed for the same initial conditions.

\begin{figure*}
 \centerline{
  \hbox{\includegraphics [   width=82mm]{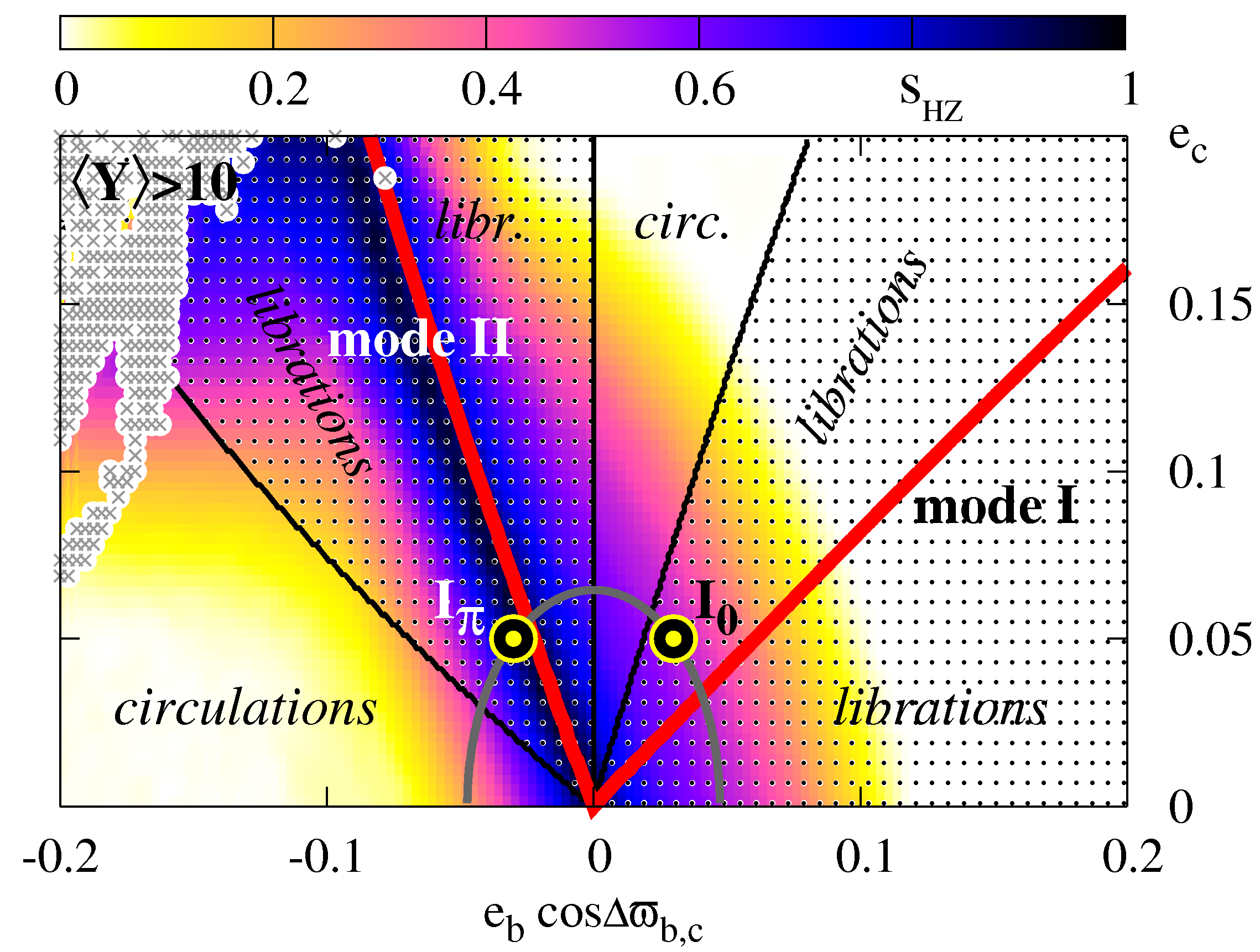}\hskip5mm
        \includegraphics [   width=70mm]{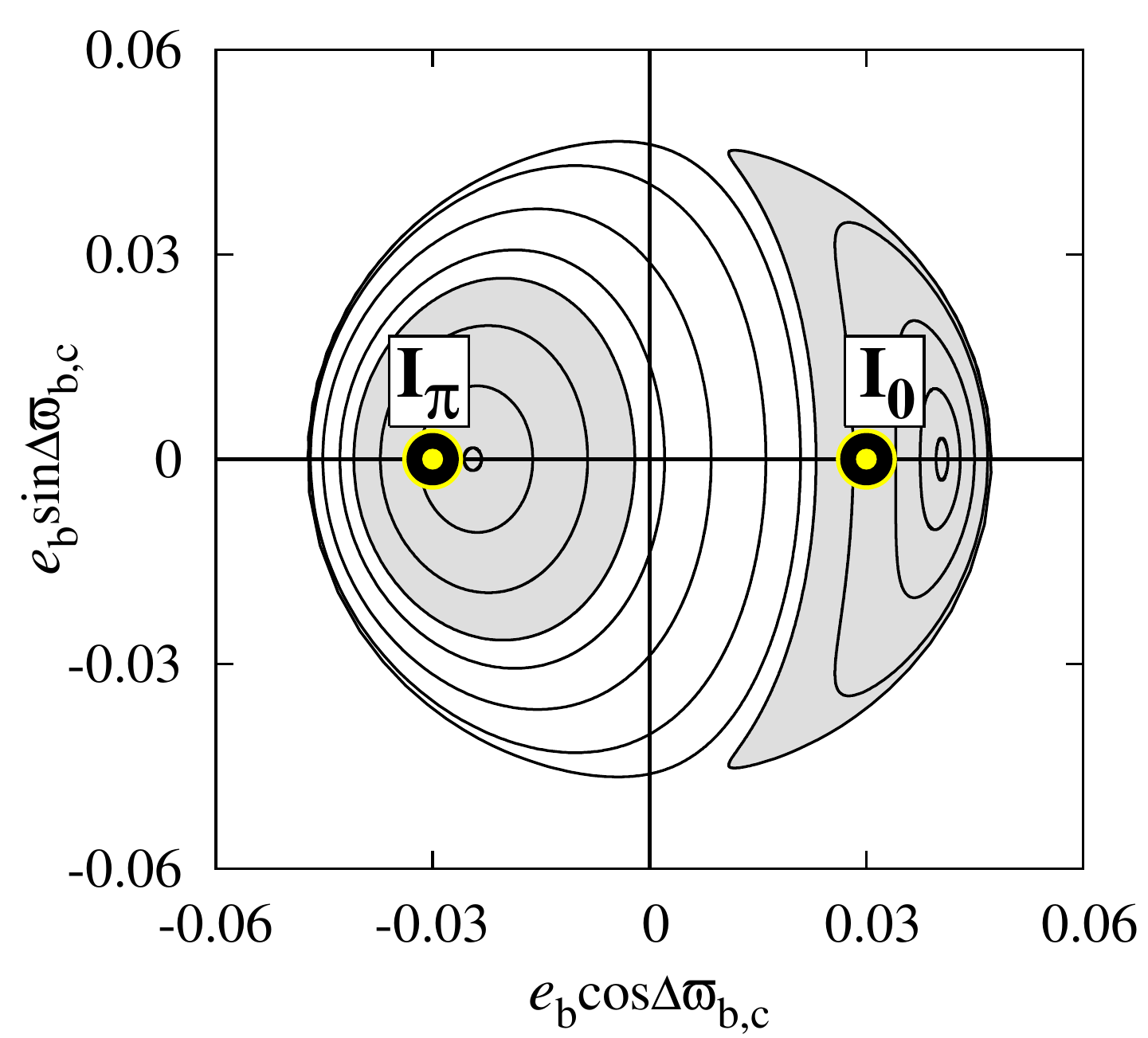}} 
 }
 \caption{
 {\em The left-hand panel.} Color coded $\sHz$ in the ${\cal S}$-plane for
$a_{\idm{b}} = 2.3$~au,  $a_{\idm{c}} = 4.6$~au (model $I_{0,\pi}$).  Gray
crosses indicate unstable configurations of Jupiters in terms of $\Ym>10$.  
Nominal elements for $\Delta\varpi_{\idm{b,c}}=0,\pi$ are labeled with $I_0$ and
$I_\pi$, respectively. Dotted areas are for initial conditions implying
librations of $\Delta\varpi_{\idm{b,c}}$ around 0 (the right-hand half-plane of
${\cal S}$) and around $\pi$ (the left-hand half-plane of ${\cal S}$). {\em The
right-hand panel.} The phase diagram calculated for a constant level of the
total angular momentum (marked with gray curve in the ${\cal S}$-plane). }
\label{fig:fig7}
\end{figure*}
Clearly, the region of large $\sHz$ lies in the neighborhood of stationary
mode~II. On contrary, the secular resonance in the configurations selected in
the positive half-plane ($\Delta\varpi_{\idm{b,c}} = 0$) preserves stable orbits
in a very small part of the HZ. This seems a general feature of the system, and
in fact it can be explained through the fundamental frequencies. Close to the
equilibria, $\Delta\varpi_{\idm{b,c}}$ oscillates around 0, or $\pi$.  To the
first significant octupole terms of ${\cal H}_{\idm{sec}}$,
\[
 \frac{\mbox{d}\,e_0}{\mbox{d}\,t} \sim A \sin\Delta\varpi_\idm{{0,b}}
 + B \sin\Delta\varpi_\idm{{0,c}}, \quad
\]
where $A$ and $B$ are coefficients dependent on the orbital elements, having the
same sign. Hence, the anti-aligned orbits of the giant planets favor small (or
much smaller) excitation of $e_0$ than the aligned jovian orbits.

The results of the next simulation are shown in Fig.~\ref{fig:fig8}. Now, we
changed the semi-major axes to $a_{\idm{b}}=2.5$~au,  $a_{\idm{c}}=4.1$~au  (our
model $V_0$, in the 2c:1b~MMR). We recall that the middle- and the right-hand
panels of Fig.~\ref{fig:fig2} are for the dynamical maps of this configuration
in the $(e_{\idm{b}},e_{\idm{c}})$-, and $(a_{\idm{b}},e_{\idm{b}})$-planes.
Figure \ref{fig:fig8} shows the  results in terms of color-coded $\sHz$ plotted
in the right-hand  half-plane of ${\cal S}$ (there is no equivalent of a stable
region in the left-hand half-plane of ${\cal S}$). We compute the $\sHz$ for
jovian orbits stable at least over $\sim 10$~Myrs. There is a perfect match of
the border of stable zone derived through the long-term integrations, and with 
the MEGNO map (both maps are plotted in Fig.~\ref{fig:fig8}, see also
Fig.~\ref{fig:fig2}), although the indicator was calculated over $\sim 0.4$~Myr
``only''. Comparing these two maps, we found a transient zone around
$(a_{\idm{b}}\cos\Delta\varpi_{\idm{b}},e_{\idm{b}}) \sim
(0.45~\mbox{au},0.05)$, in which MEGNO already detects chaotic motions, while
the system is disrupted after a few Myrs. It confirms a precise calibration of
the integration time of $\Ym$.
\begin{figure}
 \centerline{
 \includegraphics [   width=84mm]{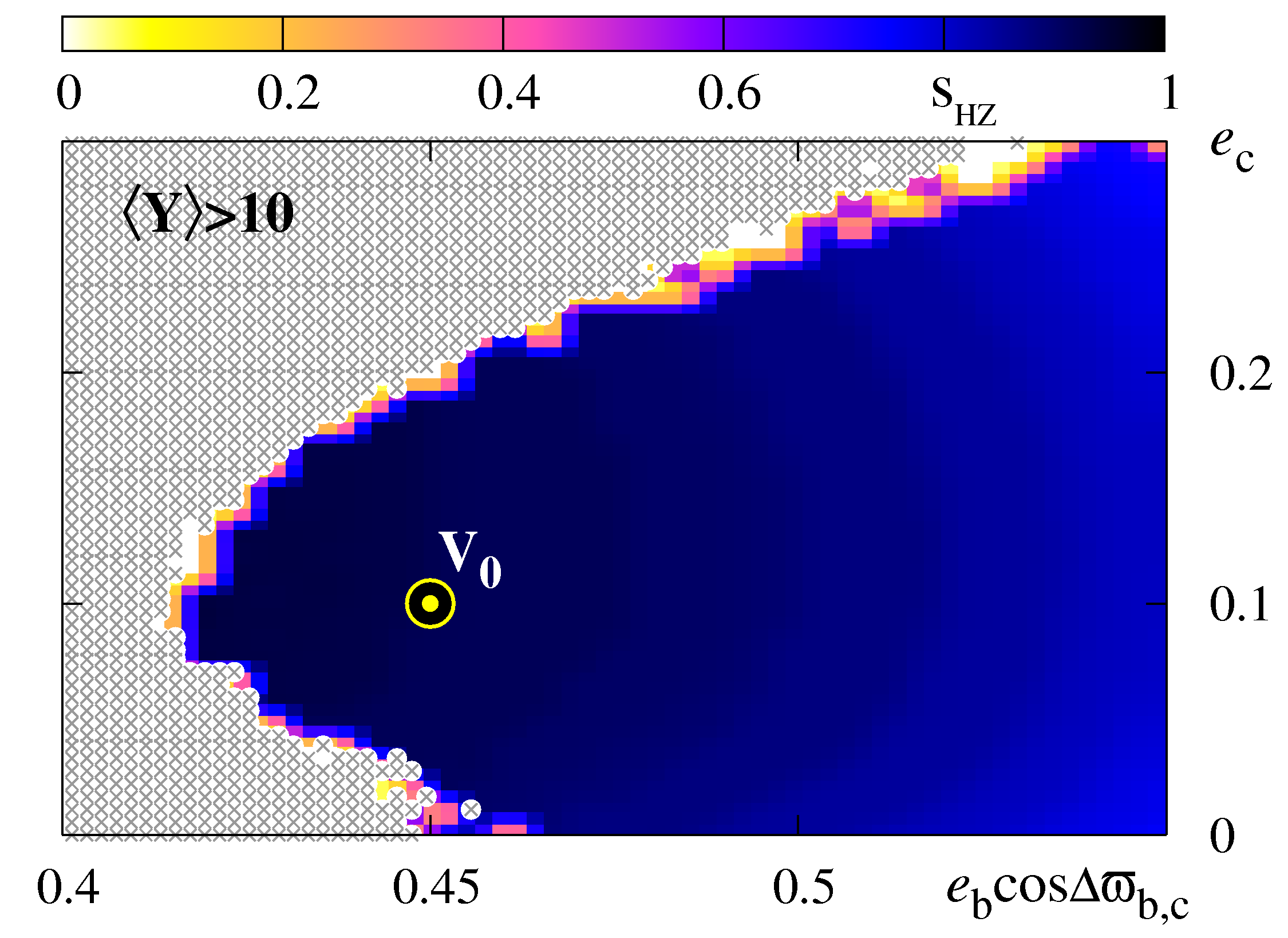} 
 }
 \caption{
 Color coded coefficient of habitability,  $\sHz$, illustrated in the  ${\cal 
 S}$-plane.  Jovian semi-major axes are $a_{\idm{b}} = 2.5$~au, and 
 $a_{\idm{c}} = 4.1$~au,  $\Delta\varpi_{\idm{b,c}}=0$ (model $V_0$, planets
 involved in the 2c:1b~MMR). Gray crosses indicate strongly unstable
 configurations of the jovian sub-system, with $\Ym>10$. 
 The tested configuration (model $V_0$, see Table~\ref{tab:tab1}) is
 marked with a circle.
}
\label{fig:fig8}
\end{figure}
%
\subsection{The habitable zone in the $\pmb{(a_{\idm{b}}, a_{\idm{c}})}$-plane}
%
Finally, we computed $\sHz$ in the $(a_{\idm{b}}, a_{\idm{c}})$-plane, fixing
the reference values of $e_{\idm{b}}=0.03$, $e_{\idm{c}}=0.05$  (our model
$I$).  For a comparison, we choose two values of
$\Delta\varpi_{\idm{b,c}}=0,\pi$. The right-hand panel of Fig.~\ref{fig:fig9} is
for initially aligned orbits of Jupiters,  the left-hand panel is for initially
anti-aligned orbits. The position of the nominal system in the parameter space
is marked with filled circles. Crosses, along straight lines, indicate unstable
solutions detected with the help of MEGNO and, in fact, they are related to the
relevant, low-order MMRs, e.g., 5c:2b, 3c:1b and 7c:2b.  Clearly, the jovian
system would be more ``friendly'' for  inhabitants of Tellus, when the orbits
are anti-aligned in the mean, and this result supports a similar conclusion
following simulations illustrated in Figs.~\ref{fig:fig4}--\ref{fig:fig7}.  The
secular $\nu_1$ resonance manifests  itself as an upside-down V-like structure
(marked with yellow color) in the right-hand panel of Fig.~\ref{fig:fig9}. It is
located between the 5c:2b and  3c:1b MMRs.  Still, even for 
$\Delta\varpi_{\idm{b,c}}=0$, there is a large region above the 3c:1b MMR line,
which is accessible for the HZ-stable orbits of Tellus.   These results are in
accord with shaded areas illustrated in Fig.~1 of \cite{Malhotra2008}.   Here,
we found that mode~II configurations (characterized by librations of
$\Delta\varpi_{\idm{b,c}}$ around $\pi$ for the neighboring trajectories), 
permit stable orbits of Tellus for almost entire observationally determined
range of semi-major axes and eccentricities implying dynamically stable
configurations of the jovian sub-system. 
\begin{figure}
 \centerline{
  \includegraphics [   width=84mm]{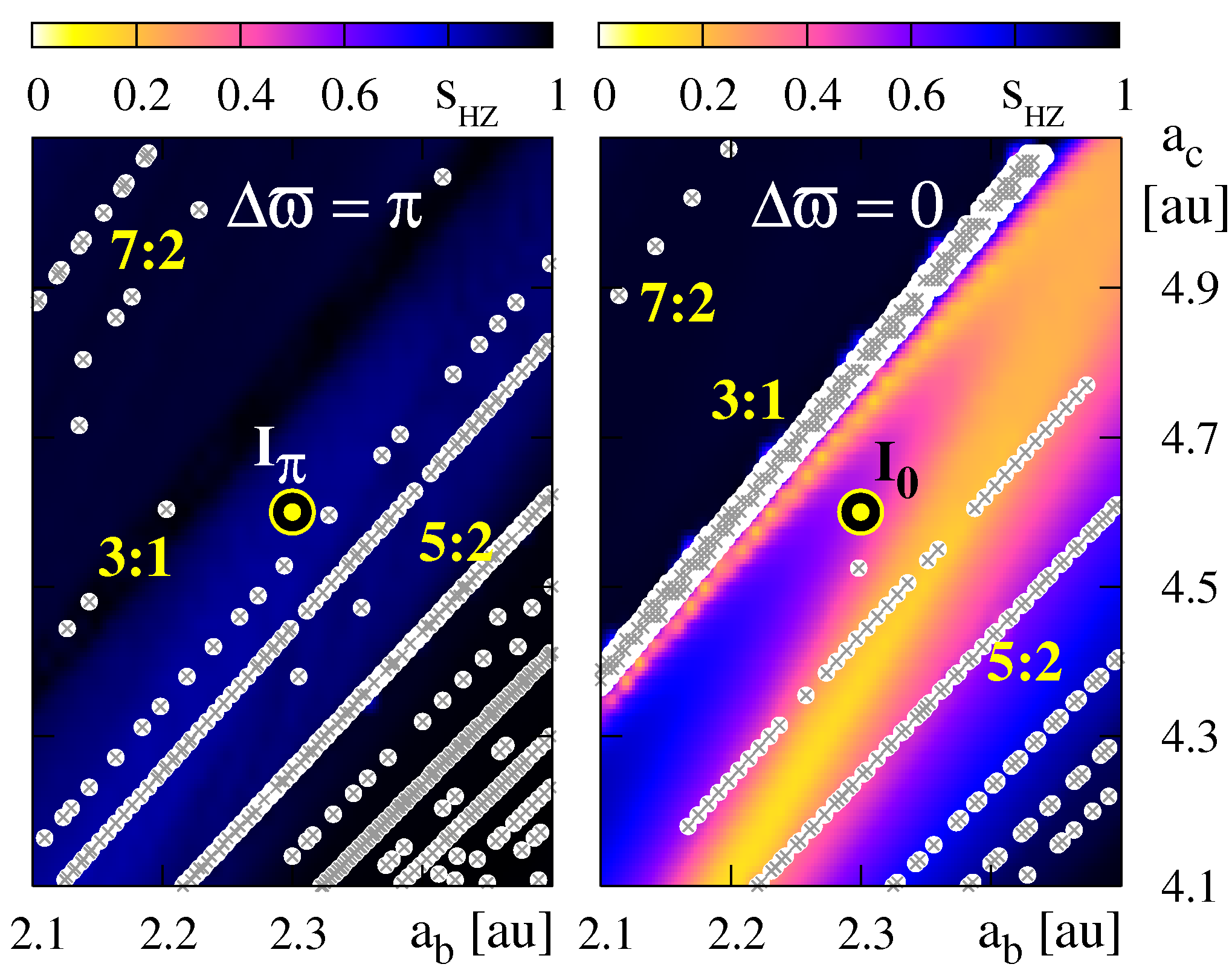} 
 }
 \caption{
 Color coded coefficient of habitability, $\sHz$, illustrated in the
 $(a_{\idm{b}}, a_{\idm{c}})$-plane. The left-hand panel is for
 $\Delta\varpi_{\idm{b,c}}=\pi$, and the right-hand panel is for
 $\Delta\varpi_{\idm{b,c}}=0$. Eccentricities of the jovian planets are
 $e_{\idm{b}}=0.03, e_{\idm{c}}=0.05$ (models $I_{0,\pi}$). Gray crosses
 indicate chaotic configurations related to MMRs (labeled) of  the b-c pair,
 $\Ym>10$.
}
\label{fig:fig9}
\end{figure}
%
\section{Conclusions} 
%
The detection levels of extrasolar planets reach  masses of super-Earths (i.e.,
Neptune-mass objects). The results of recent RV and transit surveys
[\cite{Bouchy2008,Bakos2009}, Udry 2008 (an invited talk in Toru\'n conference
{\em Extrasolar Planets in Multi-body Systems})] suggest that such small planets
are common. Hence, we expect a growing interest in dynamical studies of
extrasolar systems with terrestrial planets. Although we focus on the best
analog of the Solar system \citep{Gaudi2008}, in fact, our work concerns a whole
class of multi-planet systems involving a hypothetical Earth-like object in the
HZ of a low-mass star hosting also large, jovian planets.  To investigate the
structure of the HZ, we adopted a simple quasi-analytic theory applicable to
hierarchical configurations of the terrestrial planet and jovian companions.
Then the averaging principle can be applied, and the four-body problem may be
further simplified in terms of the restricted model. Our quasi-analytic approach
makes it possible to investigate the secular motion in an uniform way. Both the
resonant and non-resonant orbits of primary bodies (Jupiter-mass planets) may be
averaged out numerically. The secular equations of motion derived from the
analytic theory, make it possible to reduce the CPU time by a few orders of
magnitude. Hence, the method is useful to review the global features of the
secular dynamics. In accord with the assumptions of the averaging principle, the
quasi-analytic model can be also easily adopted to study long-term evolution of
similar, dynamically scaled configurations, for instance  the motion of jovian
planets in multiple systems with brown dwarfs or sub-stellar companions.
 
Regarding the \ogle{} system, we found that the structure of its HZ strongly
depends on orbits of Jupiters, and not only on their semi-major axes and
eccentricities, but also on the orbital phase and a libration mode of apsidal
angle $\Delta\varpi_{\idm{b,c}}$. From the recent theory we know
\citep{Michtchenko2004}, that apsidal librations around $0$ or $\pi$ are generic
orbital states of non-resonant, two-planet systems. In general, we found that 
the  anti-aligned mode of the jovian orbits favors much larger fraction of
stable orbits remaining entirely in the annulus of HZ between  orbital radii of
0.25~an and 0.36~au, than are permitted by the  aligned configurations. Having
in mind unconstrained parameters of the \ogle{} system, many  scenarios are
possible, even assuming that only one terrestrial planet resides in the HZ. The
disastrous $\nu_1$ secular resonance can be moved out of the HZ, through
perturbations of additional smaller planets  \citep{Malhotra2008}.   They have
shown that if such planets  exist in the system then the secular dynamics of
Earth-like planets would be very complex. Here, we also demonstrate that the
dynamics are rich in the realm of the restricted four-body problem. 

Still, there remains an open question whether the creation of Earth-like bodies
is possible in the \ogle{} system. We did preliminary simulations (Musieli\'nski
et al., in preparation) with the help of MERCURY code \citep{Chambers1999},
setting orbital parameters of the Jovian system consistent with the error bounds
\citep{Gaudi2008}. We applied the common model of elastic coagulation of small
protoplanets and planetesimals \citep[see, e.g.,][]{Raymond2008}. The initial
distribution of planetary embryos involved $\sim 50$ Moon-sized objects
distributed between 0.2~au and 2~au.  Curiously, as a typical outcome from 50
runs of 50~Myrs each, we obtained {\em single} 0.3--0.8 Earth-mass planet around
0.3~au and (sometimes) a second object of a similar mass beyond 0.6--0.8~au. The
eccentricity distribution in the small sample of runs is basically uniform, with
a number of quasi-circular orbits. These results are encouraging but they must
be confirmed by new, intensive simulations. Nevertheless, we already found some
evidence, that the terrestrial planets may emerge in the HZ of the \ogle{}
system, and we have got a new motivation to investigate the effects of  the
$\nu_1$ resonance in this region.

In this work, we applied a few different techniques to study the dynamics of
planetary systems. In particular, we apply {\em the fast indicator} (the 
symplectic MEGNO algorithm) to map the phase space of the system and to detect
chaotic (unstable) configurations.  The {\em numerical integrations} of the
equations of motion are necessary to calibrate the integration time of the fast
indicator and to test the quality of {\em  analytic expansions} with the hep of
{\em  semi-analytical averaging}. To study the structure of the HZ, it is
necessary to analyse simultaneously the short-term and the long-term  behavior
of the system.  As we think, our study  benefits from the synthesis of 
analytical and numerical techniques. 
%
\section*{Acknowledgments}
%
We thank an anonymous referee for a review and comments that improved the
manuscript. This work is supported by the Polish Ministry of Science and
Education, Grant No. 1P03D-021-29. C.M. is also supported by Nicolaus Copernicus
University Grant No.~408A. Astronomical research at the Armagh Observatory is
funded by the Northern Ireland Department of Culture, Arts and Leisure (DCAL).
\bibliographystyle{mn2e}
\bibliography{ms}
\label{lastpage}
\end{document}